\documentclass[apj,numberedappendix]{emulateapj}

\newcommand{\Linf}{L$\stackrel{\textstyle{\infty}}{~}$}
\newcommand{\sps}[1]{\mbox{$^{\mathrm{#1}}$}}
\newcommand{\sbs}[1]{\mbox{$_{\mathrm{#1}}$}}
\usepackage%
[%
pdftex,       
pagebackref, 
colorlinks=true,
linkcolor=red,
filecolor=red,
citecolor=blue,
urlcolor=magenta,
pdftitle={Models of Jupiter's Growth Incorporating Thermal and Hydrodynamic Constraints},
pdfauthor={Lissauer, Hubickyj, D'Angelo, Bodenheimer},
pdfsubject={Models of Jupiter's Growth},
pdfkeywords={planetary formation; planet-disk interaction},
bookmarks=true,
breaklinks=true,
hypertexnames=false,
pdfstartview=FitV,
hyperfootnotes=false,
bookmarksopen=false,
pdfpagemode=None]{hyperref}
\shorttitle{\textsc{Models of Jupiter's Growth}}
\shortauthors{\textsc{Lissauer, Hubickyj, D'Angelo, \& Bodenheimer}}
\begin{document}

\title{Models of Jupiter's Growth Incorporating Thermal and Hydrodynamic Constraints\altaffilmark{\dag}}
\author{Jack J. Lissauer, Olenka Hubickyj\altaffilmark{1}, Gennaro D'Angelo\altaffilmark{2}}
\affil{NASA Ames Research Center, Space Science and Astrobiology Division,
         MS~245-3, Moffett Field, CA 94035, USA}
\email{Jack.Lissauer@nasa.gov}
\email{hubickyj@pollack.arc.nasa.gov}
\email{gennaro.dangelo@nasa.gov}
\and
\author{Peter Bodenheimer}
\affil{UCO/Lick Observatory, Department of Astronomy and
         Astrophysics, University of California, Santa Cruz, 
         CA 95064, USA}
\email{peter@ucolick.org}
\slugcomment{\today}
\altaffiltext{1}{Also at UCO/Lick Observatory, University of California, Santa Cruz.}
\altaffiltext{2}{NASA Postdoctoral Fellow.}
\altaffiltext{$^\dag$}{%
                      To appear in the journal %
                      \textsc{Icarus}. %
                                        }
\begin{abstract}
We model the growth of Jupiter via core nucleated accretion, 
applying constraints from hydrodynamical processes that result 
from the disk--planet interaction. We compute the planet's 
internal structure using a well tested planetary formation code 
that is based upon a Henyey-type stellar evolution code. 
The planet's interactions with the protoplanetary disk are 
calculated using 3-D hydrodynamic simulations.  
Previous models of Jupiter's growth have taken the radius of 
the planet to be approximately one Hill sphere radius,
$R_{\mathrm{H}}$. However, 3-D hydrodynamic simulations show 
that only gas within $\sim 0.25\,R_{\mathrm{H}}$ remains bound 
to the planet, with the more distant gas eventually participating in the 
shear flow of the protoplanetary disk. Therefore in our new 
simulations, the planet's outer boundary is placed at the location 
where 
gas has the thermal energy to reach the portion of the flow not
bound to the planet.  
We find that the smaller radius increases the time required for 
planetary growth by $\sim 5\%$. Thermal pressure limits the 
rate at which a planet less than a few dozen times as massive 
as Earth can accumulate gas from the protoplanetary disk, 
whereas hydrodynamics regulates the growth rate for more massive 
planets. 
Within a moderately viscous disk, the accretion rate peaks when 
the planet's mass is about equal to the mass of Saturn. In a
less viscous disk hydrodynamical limits to accretion are
smaller, and the accretion rate peaks at lower mass.  
Observations suggest that the typical lifetime of massive disks 
around young stellar objects is $\sim 3\,\mathrm{Myr}$. 
To account for the dissipation of such disks, we perform some 
of our simulations of Jupiter's growth within a disk whose 
surface gas density decreases on this timescale.
In all of the cases that we simulate, the planet's effective 
radiating temperature rises to well above $1000\,\mathrm{K}$ 
soon after hydrodynamic limits begin to control the rate of gas 
accretion and the planet's distended envelope begins to contract.
According to our simulations, proto-Jupiter's distended and 
thermally-supported envelope
was too small to capture the planet's current retinue of irregular 
satellites as advocated by Pollack et al.\ 
[Pollack, J.B., Burns, J.A., Tauber, M.E., 1979. Icarus 37, 587--611].
\end{abstract}

\section{Introduction}
\label{sec:introduction}
\defcitealias{bodenheimer1986}{BP86}
\defcitealias{bodenheimer2000b}{BHL00}
\defcitealias{gennaro2003a}{DKH03}
\defcitealias{hubickyj2005}{HBL05}
\defcitealias{pollack1996}{PHBLPG96}
According to the core nucleated accretion model, giant planets 
begin their growth via the same process of agglomeration of 
solid bodies as do terrestrial planets; however, unlike 
terrestrials, the solid cores of giant planets reach masses 
large enough to capture substantial amounts of gas from their 
star's protoplanetary disk before said disk dissipates 
\citep{LS2007}.
Previous models of this process have simulated either the thermal 
factors that limit the ability of a planet to retain gas
(\citealp{bodenheimer1986}, hereafter \citetalias{bodenheimer1986};
\citealp{pollack1996}, hereafter \citetalias{pollack1996};
\citealp{bodenheimer2000b}, hereafter \citetalias{bodenheimer2000b};
\citealp{ikoma2000,hubickyj2005}, hereafter \citetalias{hubickyj2005};
\citealp{alibert2005a,alibert2005b,marley2007})
or the disk interaction physics that governs the flow of gas to 
a planet
(\citealp{rnelson2000}; \citealp{gennaro2003a}, hereafter 
\citetalias{gennaro2003a}; \citealp{bate2003}).  
Here we consider both thermal and gas flow limits to giant 
planet growth, and present the first models of the growth of 
Jupiter that are constrained by detailed simulations of both 
of these factors.

A planet of order one to several Earth masses (M$_{\oplus}$) 
at a distance of about $5\,\mathrm{AU}$ from the central star
is able to capture an atmosphere from the protoplanetary disk 
because the escape speed from its surface is large compared to 
the thermal velocity of gas in the disk. However, such an 
atmosphere is very tenuous and distended, with thermal pressure 
pushing gas  outwards and thereby limiting further accretion of 
gas. The key factor governing the ability of planet to 
accumulate additional gas when the mass of the atmosphere is 
less than the mass of the core is the planet's ability to radiate 
the energy that is provided to it by the accretion of 
planetesimals and gravitationally-induced compression of gas. 
The escape of this energy cools  the gaseous  envelope, allowing 
it to shrink and thereby enabling  more gas to enter the planet's 
gravitational domain. Evolution occurs slowly, and hydrostatic 
structure is generally a very good approximation. 
Once a planet has enough mass for its self-gravity to compress 
the envelope substantially, its ability to accrete additional 
gas is limited only by the amount of gas available.  
Hydrodynamic limits allow quite rapid gas flow to a planet in 
an unperturbed disk.  But a planet alters the disk by accreting
material from it and by exerting gravitational torques upon it 
\citep{lin1979,goldreich1980}.
Both of these processes can lead to gap formation and isolation 
of the planet from the surrounding gas. 

Our approach is to follow the physical structure and thermal 
evolution of the growing giant planet in the spherically 
symmetric (one-dimensional) quasi-hydrostatic approximation, 
and  to incorporate the three-dimensional hydrodynamic 
interactions between the planet and the circumstellar disk via 
boundary conditions at the planet's outer `surface'.
Mass and energy transport within the planet are followed using 
the same planetary evolution code that we have
employed in previous models of giant planet formation 
(\citetalias{bodenheimer1986,pollack1996,bodenheimer2000b,%
hubickyj2005}; \citealp{marley2007}).

\citet{bodenheimer1986} prescribed the accretion rate of solids 
to be constant with time. \citet{pollack1996} replaced this model 
by assuming that the planet was an isolated embryo that underwent 
runaway growth within a disk of dynamically cold, non-migrating, 
planetesimals.
The accretion rate of solids depends upon the distribution of 
planetesimals as well as the planet's mass and its effective 
radius for accretion of planetesimals. The planet's capture
cross-section was computed using the physical properties of 
the planet determined by the planetary structure calculation. 
The rate at which the planet 
accreted solids, $\dot{M}_Z$, for specified planet
cross-section and disk surface density, eccentricities 
and inclinations of 
planetesimals within the planet's feeding zone, was determined 
using formulae that \citet{greenzweig1992} derived from 3-body 
numerical studies of planetesimal trajectories. 
This prescription has been used with slight modifications in 
most of our subsequent calculations, including all of those 
presented herein.   

Our previous simulations have used simple \textit{ad hoc} 
prescriptions
for the interactions of the planet with the gaseous disk.  
We placed the outer boundary of the 
planet near its Hill sphere radius, $R_{\mathrm{H}}$, during 
most of its growth.  
The radius of the planet's Hill sphere is given by:
\begin{equation}
R_{\mathrm{H}}=r_p%
\left(\frac{M_p}{3\,M_{\star}}\right)^{\frac{1}{3}},
\label{eq:RH}
\end{equation}
where $M_p$ ($=M_{XY}+M_Z$) is the (gas + solids) mass 
of the planet, $M_{\star}$ 
the mass of the star, and $r_p$ is the orbital radius 
of the planet. 
More precisely, \citetalias{bodenheimer2000b}, 
\citetalias{hubickyj2005}, and \citet{marley2007}
took the planet's boundary to be the location where the thermal 
velocity of the H$_2$ gas molecules gave them sufficient energy 
to move upwards to $1\,R_{\mathrm{H}}$ from the planet's center. 
We limited the rate at which the planet could accrete gas from 
the disk to a maximum of 
$\sim 10^{-2}\,\mathrm{M}_{\oplus}$ per year, which is  
approximately the Bondi rate. We extended many of our runs to a 
pre-determined mass limit of a Jupiter mass or more, and in a 
few cases we followed the ensuing phase of planetary
contraction for $4.5\,\mathrm{Gyr}$.  
But because of the approximate treatment of the later phases
of gas accretion, 
we have always emphasized as our primary results the crossover
time (when the planet's gas mass equals the mass of its 
condensables) and the corresponding crossover mass. The total 
formation time for the planet is generally only slightly longer 
than the crossover time.

We present herein results of new simulations using our venerable 
1-D planetary formation code to follow the evolution of the planet's 
structure, but now incorporating 3-D hydrodynamic calculations 
for prescriptions of the planet's size and maximum rates of gas 
accretion. 
In some of our calculations, we gradually reduce the 
density of gas within the surrounding disk to provide a more 
realistic simulation of the final phases of the planet's growth.

In the models presented herein, we neglect orbital migration.
During the phase of runaway gas accretion, the amount of radial
migration that is expected before the planet reaches one Jupiter-mass 
is on the order of 20\% of its initial orbital radius \citep{gennaro2008}. 
Orbital decay due to resonant torques during the phase of slow gas 
accretion (Phase~II) may be more substantial. However, a number 
of mechanisms may conspire to reduce those migration rates 
\citep[see][ for a review] {papaloizou2007}. There is presently a great 
deal of uncertainty surrounding these issues, so rather than rely on
some poorly constrained  and not yet well-understood migration 
mechanism, our simulations simply assume that the orbit of the planet 
remains fixed.  The differing migration scenarios may affect giant 
planet growth in different ways, but our assumption of no migration
is extreme in the sense that the isolation mass of a core within a 
planetesimal disk is larger for any non-zero migration of the planet, 
because the radial motion of the planet brings it into regions of the 
disk that are undepleted of planetesimals 
\citep{lissauer1993,alibert2005a}. 
So migrating planets, or planetesimals migrating as a result of
gas drag \citep{kary1993,kary1995}, are likely capable of forming 
somewhat larger cores for a given location and disk surface mass 
density of solids than are the non-migrating planets that we 
simulate herein.
Competing embryos in nearby accretion zones can act in the opposite 
sense from the above mentioned processes by removing solids from the
planet's reach. But if the planet accretes an embryo,
said embryo can bring with it solids from somewhat beyond the 
planet's nominal accretion zone.

Our 1-D accretion code is described in \citetalias{hubickyj2005} 
and references therein. 
Details on the 3-D hydrodynamic numerical code can be 
found in \citetalias{gennaro2003a} and references therein. 
We present our limits on the planet's physical extent and gas 
accretion rate, derived from 3-D hydrodynamic simulations, 
in Section~\ref{sec:envsize}. 
Section~\ref{sec:simupar} discusses the physical parameters 
for our simulations. 
The results of our calculations are presented in 
Section~\ref{sec:results}. 
The scenario of capture of irregular satellites within proto-jupiter
distended and thermally-supported envelope \citep{pollack1979} is
discussed within the framework of our models for the growth of Jupiter
in Section~\ref{sec:satellites}.
We conclude in Section~\ref{sec:summary} with a discussion 
of our findings and their implications.

\section{Envelope Size and Maximum Gas Accretion Rates}
\label{sec:envsize}
Three-dimensional simulations of a disk with an embedded 
planet are used to estimate (\textit{i}) the region of space 
within which gas is bound to a planetary core 
(Section~\ref{sec:outerboundary}) and (\textit{ii}) 
the maximum accretion rate at which the disk can feed the inner 
parts of a growing planet's Hill sphere 
(Section~\ref{sec:accretionrates}). 

\subsection{Outer Boundary of Planet's Envelope}
\label{sec:outerboundary}
In order to evaluate the volume of gas that is gravitationally
bound to a planet, we adopt disk models similar to those 
described in \citetalias{gennaro2003a}.  
The simulation region extends from $0$ to $2\pi$ in azimuth and 
over a radial range from $2$ to $13\,\mathrm{AU}$,
so that the disk boundaries are well separated from the planet's 
orbit.
The pressure scale height of the disk at the planet's orbit, 
$H_p$, is taken to be $5$\% 
of the distance to the star; this corresponds to a temperature of 
$T = 115\,\mathrm{K}$ for a gas of mean molecular weight $2.25$ 
at a distance of $5.2\,\mathrm{AU}$ from a $1\,\mathrm{M}_{\odot}$
(solar mass) star. 
The dimensionless disk viscosity parameter is assumed to be 
$\alpha=4\times 10^{-3}$. 
We consider planet masses ranging from $10\,\mathrm{M}_{\oplus}$ 
to $50\,\mathrm{M}_{\oplus}$, because at smaller masses the 
planet's envelope is very tenuous, and because at a mass 
exceeding $\sim 70\,\mathrm{M}_{\oplus}$ the planet self-compresses 
to a size much smaller than that of its Hill sphere. We use grid 
systems that resolve the mass density and the velocity field in 
the vicinity of the planet on length scales shorter than $4$\% of 
the Hill radius; thus the circumplanetary subdisk is also resolved.
Simulations are started from an unperturbed Keplerian disk, whose 
rotation is corrected for effects of the gas pressure gradient. 
Models are evolved for about $250$--$300$ orbital periods, at 
which time the flow has reached a quasi-stationary state.
Tracer (massless) particles are deployed in the quasi-stationary
flow, within a distance of approximately $R_{\mathrm{H}}$
from the core, and their trajectories are integrated for tens
of orbital periods of the planet around the Sun.
Several initial distributions of tracers are used, for purposes 
of a sensitivity study, containing from $500$ to $6000$ particles.

The tracers are advected by the flow field and a second-order
Runge-Kutta algorithm is used to advance their position in time.
Gas velocities are interpolated to the tracers' positions via
a monotonized harmonic mean \citep{gennaro2002}, which is 
second-order accurate and capable of dealing with large 
gradients and shock conditions. Therefore, the procedure is
second-order accurate in both space and time.

We adopt a conservative approach to identify trajectories
trapped inside the gravitational potential of the planet.
Indicating with $s_{i}(t)$ the distance of the $i$-th particle 
from the center of the planet at time $t$, a tracer is marked 
as bound if $s_{i}(t)/s_{i}(0)<\zeta$ along its calculated trajectory,
where $s_{i}(0)<R_{\mathrm{H}}$.
The number of bound tracers can grow as $\zeta$ increases
because particles can temporarily move farther away from their
deployment sites (i.e., $\zeta>1$) before approaching the planet's 
center and being finally accreted.
However, the number of trapped trajectories is expected to
eventually converge, upon increasing  $\zeta$, since particles that 
escape to 
librating or circulating orbits quickly move far away from the planet.
Therefore, to select bound tracers, we increase the value of 
the parameter $\zeta$
until the number of selected trajectories does not change any 
longer, at which point we assume that the number of trapped 
particles has converged. Finally, we check that tracers discarded
according to this procedure move out of the planet's Hill sphere
and thus return to the circumstellar disk.
Note that the approach adopted here does not characterize as 
bound material gas that moves along trajectories originating outside 
of the Hill sphere and accreting onto the planet.

\begin{figure*}[th!]
\centering%
\resizebox{0.60\hsize}{!}{\includegraphics[trim=0 5 0 0, clip=true]{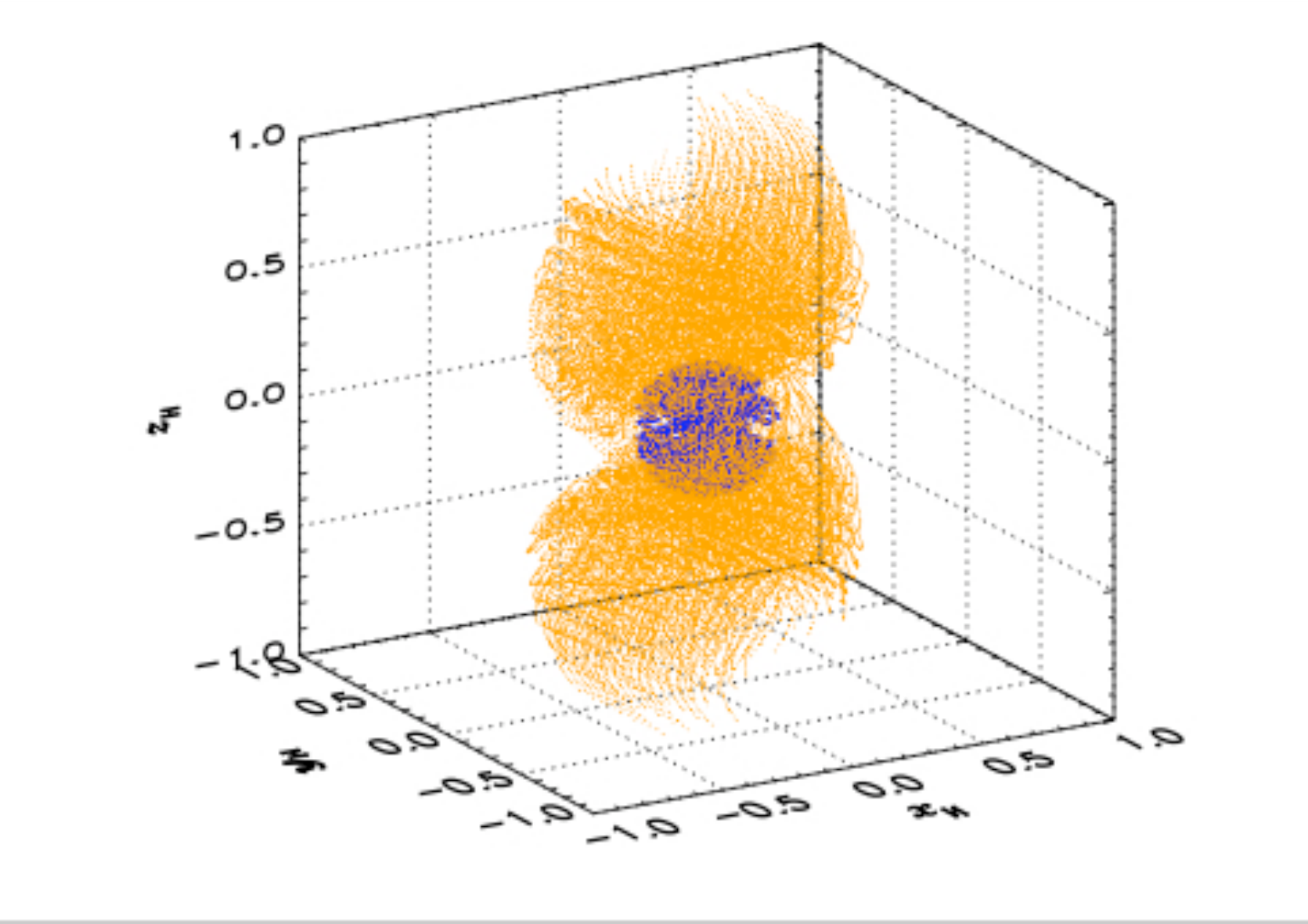}}
\resizebox{0.60\hsize}{!}{\includegraphics[trim=0 5 0 0, clip=true]{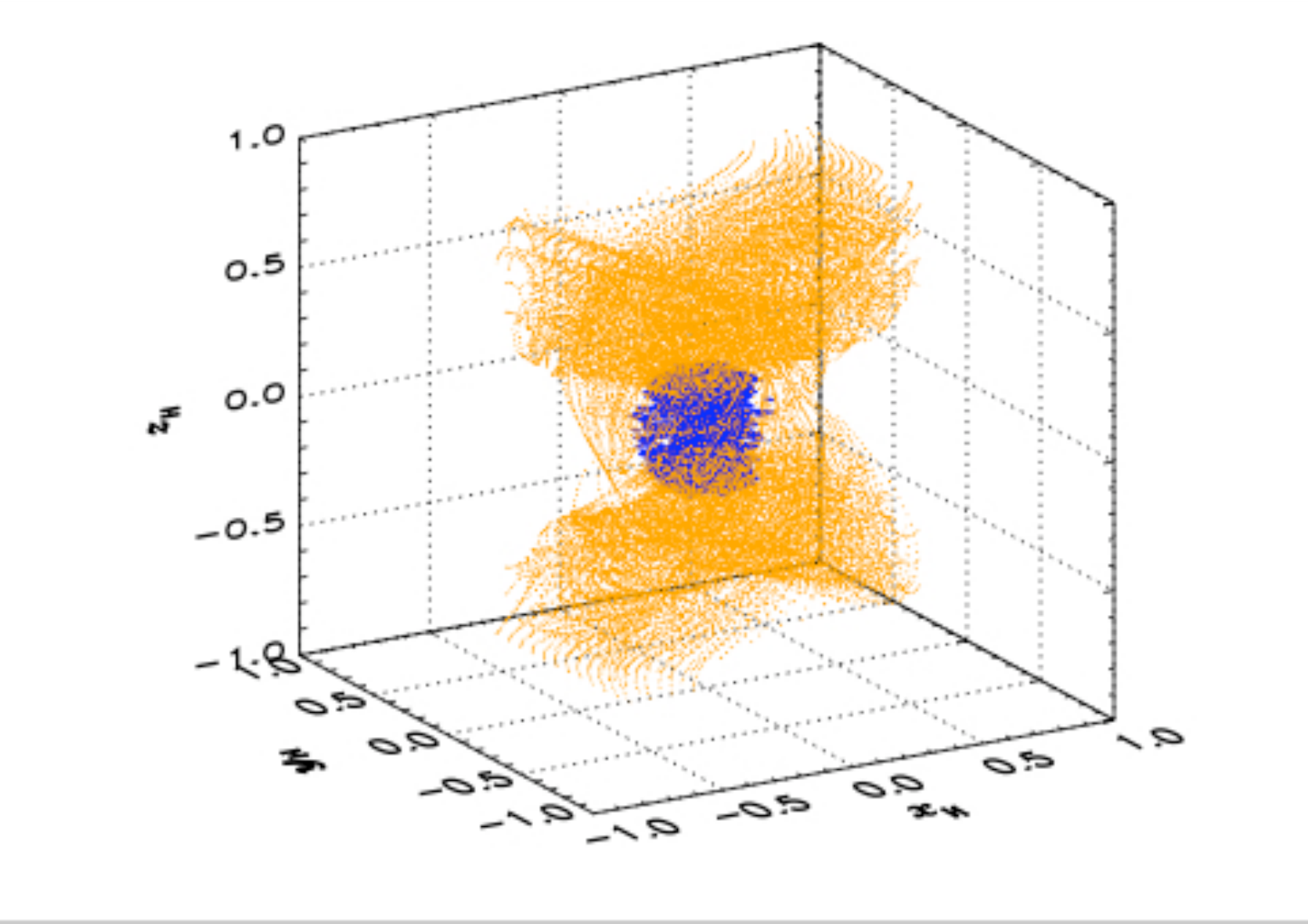}}
\caption{Trajectories in a frame rotating at the angular velocity
         of the planet (around the Sun) of $\sim 2000$ tracer 
         particles that are bound to a $10\,\mathrm{M}_{\oplus}$
         (top) and a $30\,\mathrm{M}_{\oplus}$ (bottom)
         point-mass planet, located at the origin.
         Bound particles were selected according to the procedure
         described in the text. Axes are in units of the Hill 
         radius, $R_{\mathrm{H}}$. Each dot marks the position 
         occupied by a tracer at a given time.
         Positions within $R_{\mathrm{H}}/4$ of the center of
         the planet are marked as blue dots, while positions
         beyond this distance are marked as orange dots.
         }
         \label{fig:boundtracers}
\end{figure*}
The initial distribution of bound particles is, in general, not 
spherically symmetric around the core, but it is roughly 
symmetric relative to the disk midplane, as displayed in 
Fig.~\ref{fig:boundtracers}. 
This figure shows positions (over about $3$ orbital periods of
the planet around the Sun) of tracers bound to a 
$10\,\mathrm{M}_{\oplus}$ (top) and a $30\,\mathrm{M}_{\oplus}$ 
(bottom) core and selected according to the 
procedure outlined above. The center of the planet is located at
the origin of the axes and distances are normalized to 
$R_{\mathrm{H}}$.
Blue dots indicate positions with $s_{i}(t)<R_{\mathrm{H}}/4$,
whereas orange dots mark positions of particles trapped beyond
$R_{\mathrm{H}}/4$.

\begin{figure*}
\centering%
\resizebox{0.65\hsize}{!}{\includegraphics{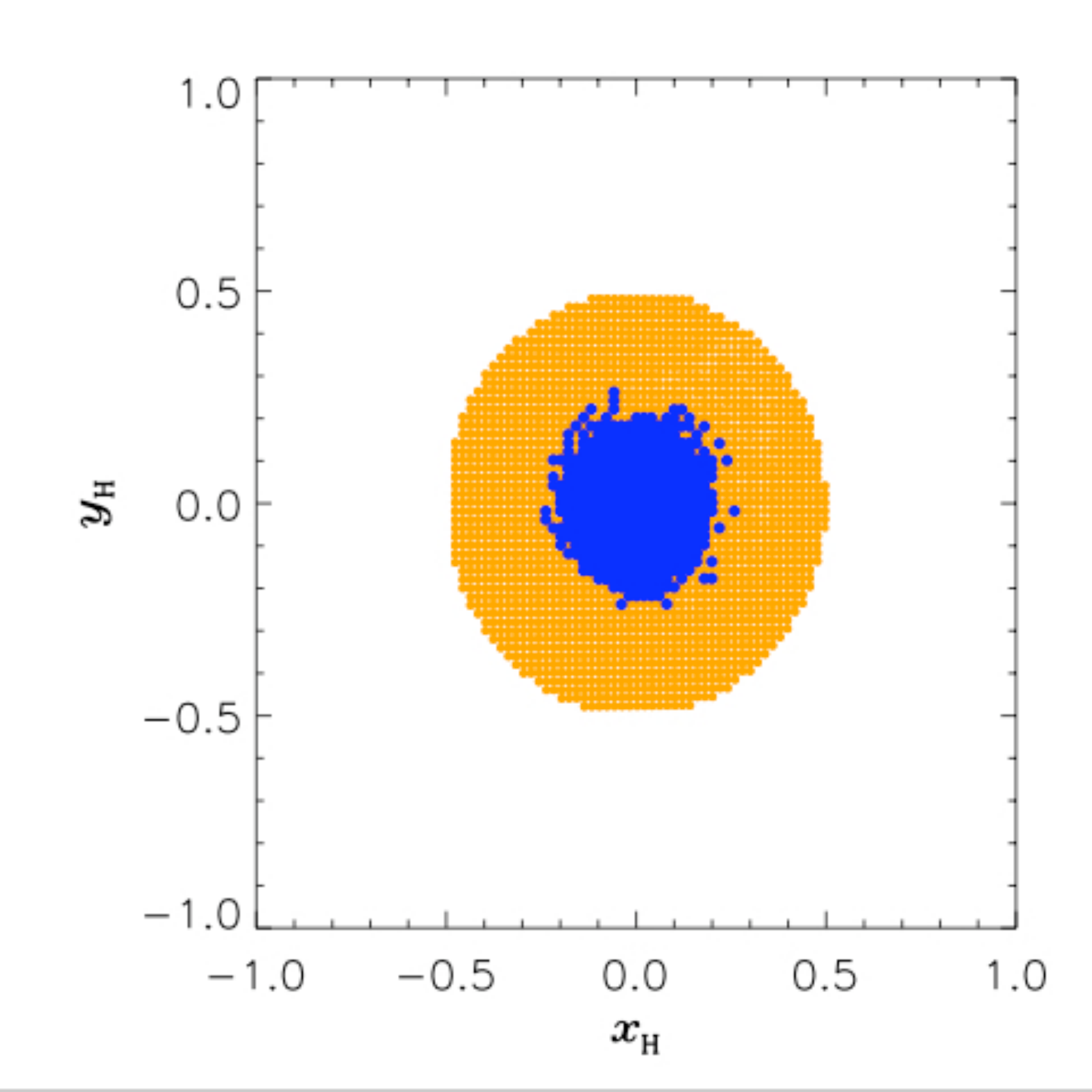}%
                      \includegraphics{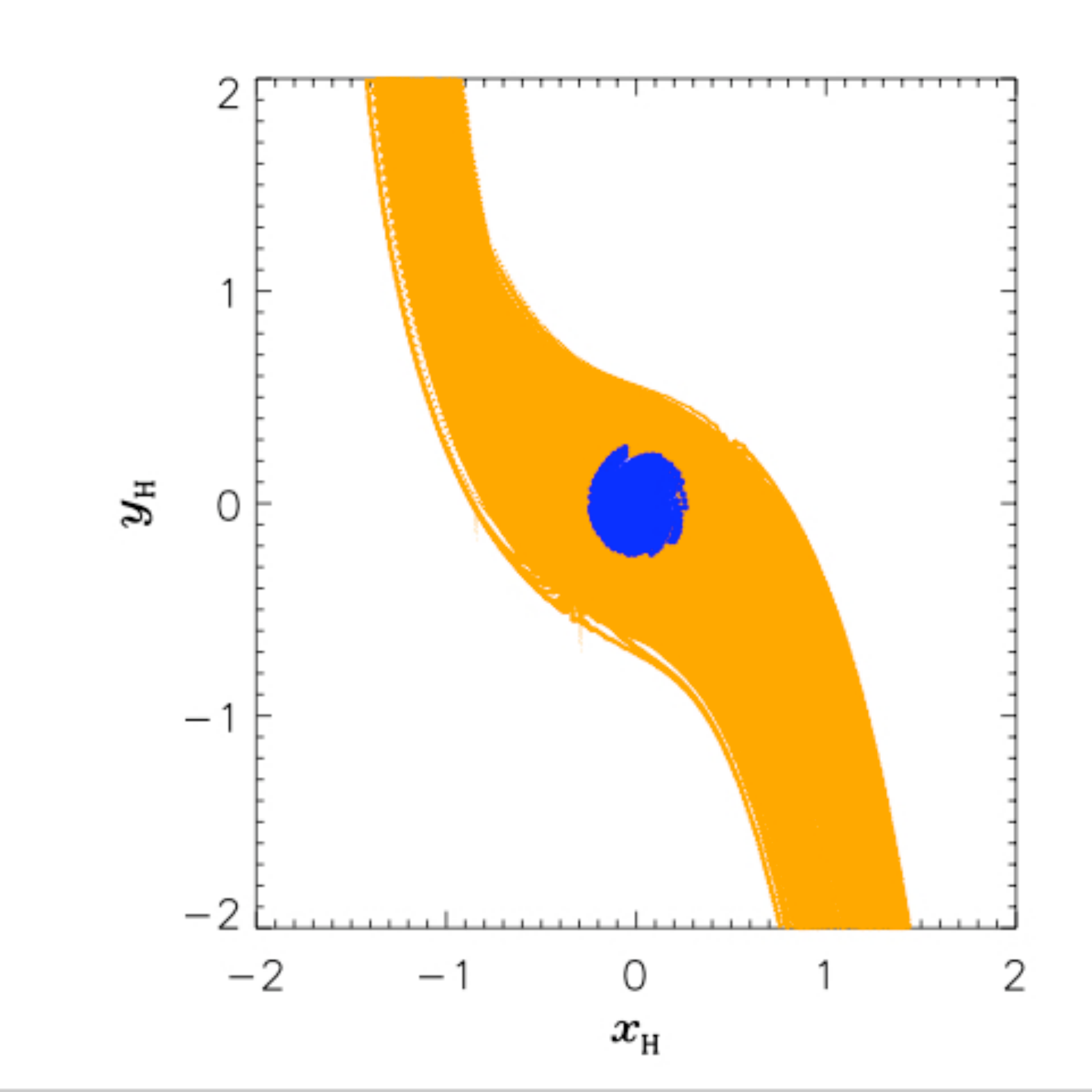}}
\resizebox{0.65\hsize}{!}{\includegraphics{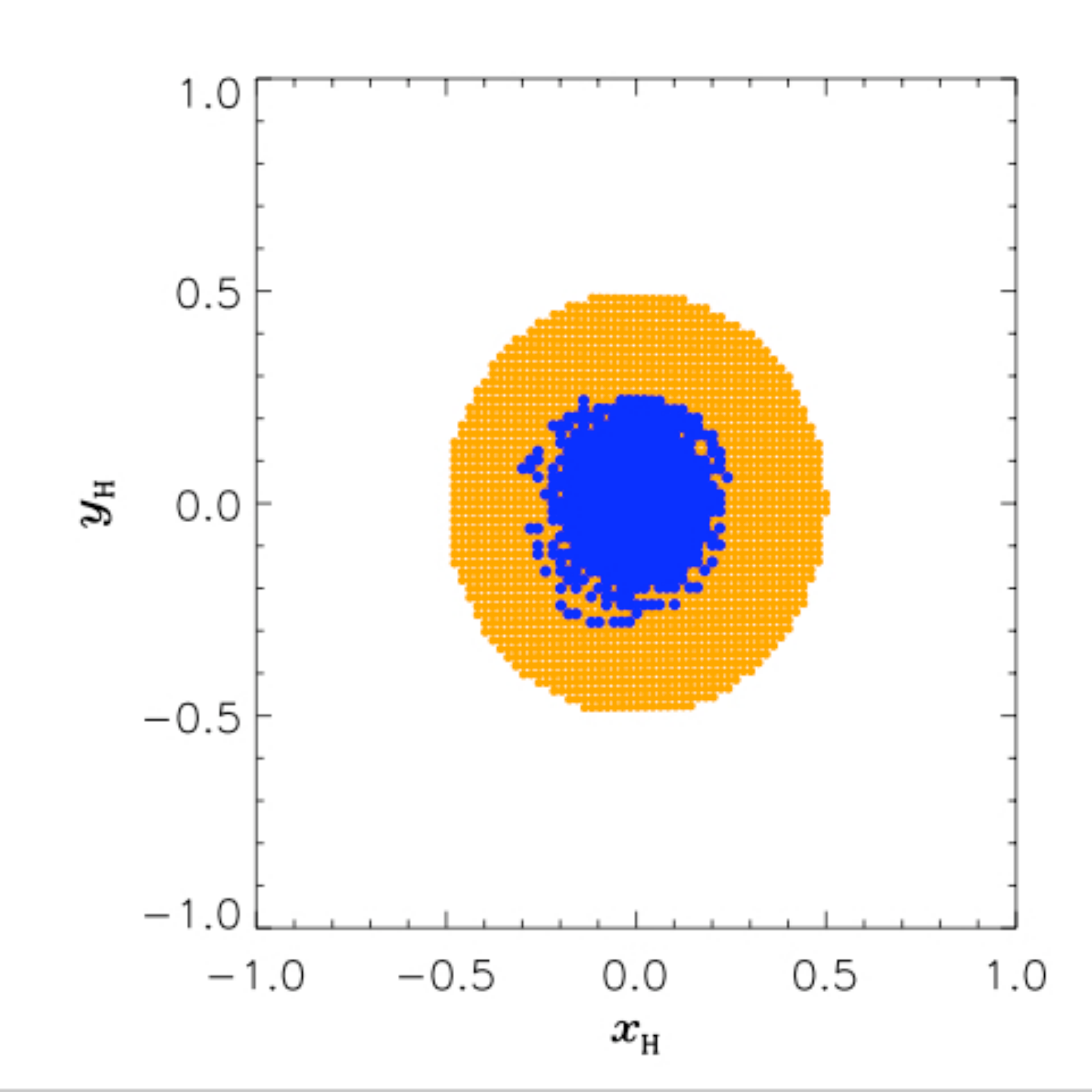}%
                      \includegraphics{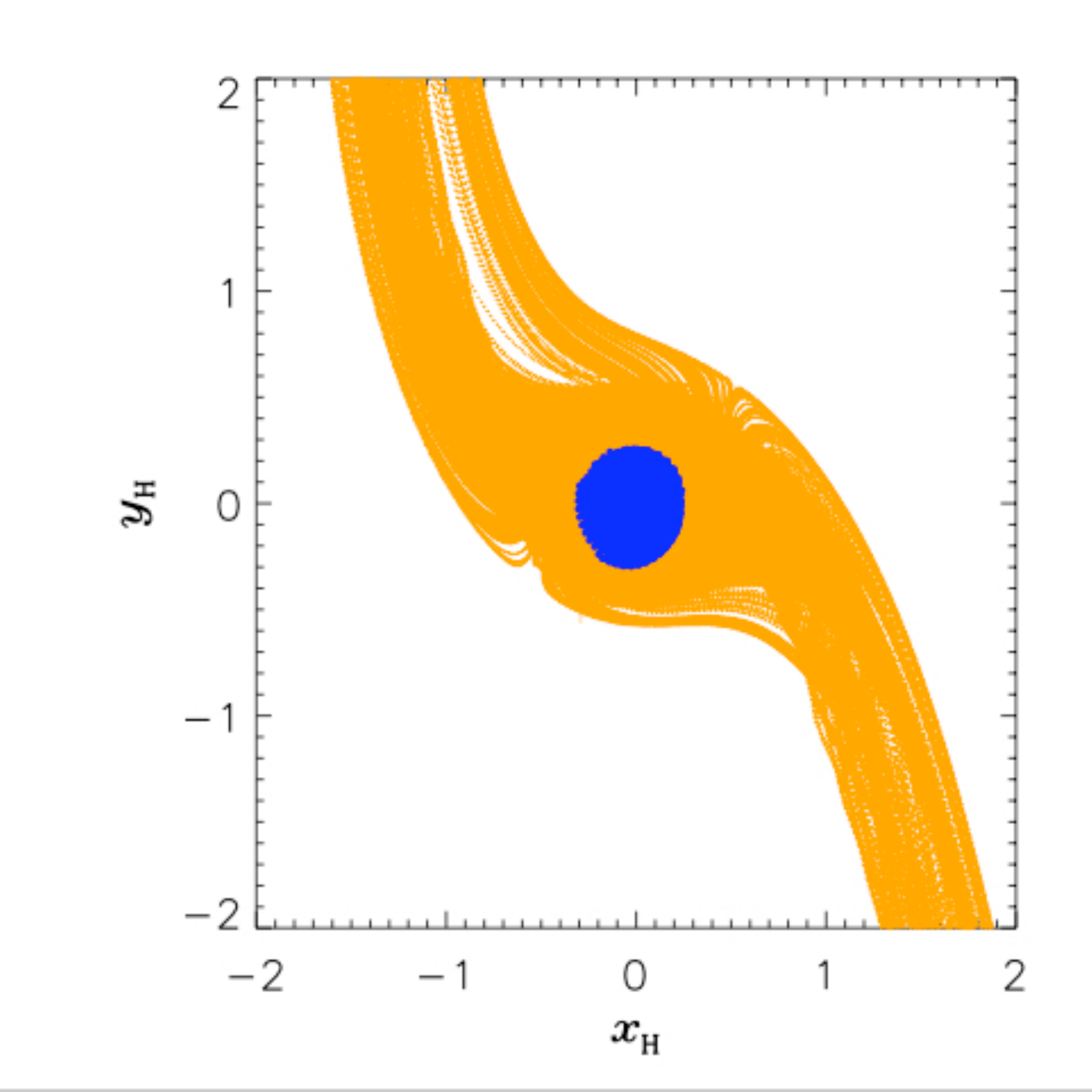}}
\caption{Initial positions (left) and trajectories (right),
         in a frame rotating at the angular velocity of the
         planet, of about $1000$ tracer particles deployed
         close to the disk midplane and within
         $R_{\mathrm{H}}/2$ of a $10\,\mathrm{M}_{\oplus}$
         (top) and a $30\,\mathrm{M}_{\oplus}$ (bottom)
         point-mass planet located at the origin.
         Blue circles (left) and dots (right) indicate bound 
         particles according to the procedure described in the
         text.
         Orange circles or dots represent particles that leave
         the planet's Hill sphere and return to the 
         circumstellar disk.
         }
         \label{fig:tracers_xy}
\end{figure*}
Given the necessity to describe the region containing bound 
particles with a single length for our 1-D planetary structure 
calculations, 
we estimate the radius of the largest sphere centered on the 
core and enclosed in the initial distribution of selected 
particles. This is done by considering the area obtained from 
the intersection of the volume occupied by selected particles 
with the disk midplane and evaluating the radius of the largest
circle centered on the core and enclosed in this area.
For the cases we investigate, we obtain a radius ranging from
$\sim R_{\mathrm{H}}/4$ to $\sim R_{\mathrm{H}}/3$.
An example is illustrated in Fig.~\ref{fig:tracers_xy}.
The left panel shows the initial positions of particles deployed
close to the disk midplane and within $R_{\mathrm{H}}/2$ of a
$10\,\mathrm{M}_{\oplus}$ (top) and a $30\,\mathrm{M}_{\oplus}$ 
(bottom) core: blue circles represent bound 
tracers whereas orange circles represent particles that 
eventually escape from the core's Hill sphere, as can be seen 
from their trajectories displayed in the right panel.

\subsection{Gas Accretion Rates}
\label{sec:accretionrates}
We use the set of simulations in \citetalias{gennaro2003a}, 
together with analogous new simulations, to evaluate the 
maximum accretion rate at which a disk, perturbed by an 
embedded planet, can deliver gas to the planet's vicinity. 
As before, we first consider disks with viscosity 
$\alpha=4\times 10^{-3}$ and local temperature 
$T = 115 \,\mathrm{K}$. 
The numerical resolution is comparable to that of simulations 
discussed in the previous section. In these calculations, 
the accretion of gas proceeds almost uninhibited and is only 
limited by tidal effects (e.g., the formation of a density
gap) or lack of supply from the disk. Therefore, accretion 
rates described in this section represent upper limits to the
rate at which the disk can feed the inner parts of a 
planet's Hill sphere.

We obtain gas accretion rates, $\dot{M}_{XY}$
($M_{XY}$ denotes the mass of the H/He component of 
the planet), for planets 
ranging in mass from about $1\,\mathrm{M}_{\oplus}$ to 
$2\,\mathrm{M}_J$ (Jupiter masses). 
Data can be well 
fitted with a second-order polynomial written as
\begin{equation}
\log{\left(\frac{\dot{M}_{XY}}{\Sigma_{\mathrm{g}}
 r^2_p/P}\right)}\approx%
c_0+c_1 \log{\left(\frac{M_p}{M_{\star}}\right)}%
  +c_2 \log^2{\left(\frac{M_p}{M_{\star}}\right)},
\label{eq:dotMxy}
\end{equation}
where $\Sigma_{\mathrm{g}}$ is the unperturbed surface density 
of gas at the orbital radius of the planet, $r_p$, 
and $P$ is the planet's orbital period. The coefficients are: 
$c_0=-18.67$, $c_1=-8.97$, and $c_2=-1.23$.
Gas accretion rates obtained from hydrodynamical models can 
be re-scaled by the initial mass density in the disk, at the 
planet's orbital radius, because continuity and momentum 
equations that are solved in the calculations 
\citep[see][]{gennaro2005} can be normalized to an initial 
mass density (at $r_p$, for example) when pressure 
is directly proportional to mass density.
Therefore, Eq.~(\ref{eq:dotMxy}) can be used to derive 
a gas accretion rate as a function of the planet's mass, its 
orbital radius, and the unperturbed surface density.

\begin{figure*}
 \centering
 \resizebox{0.65\hsize}{!}{\includegraphics{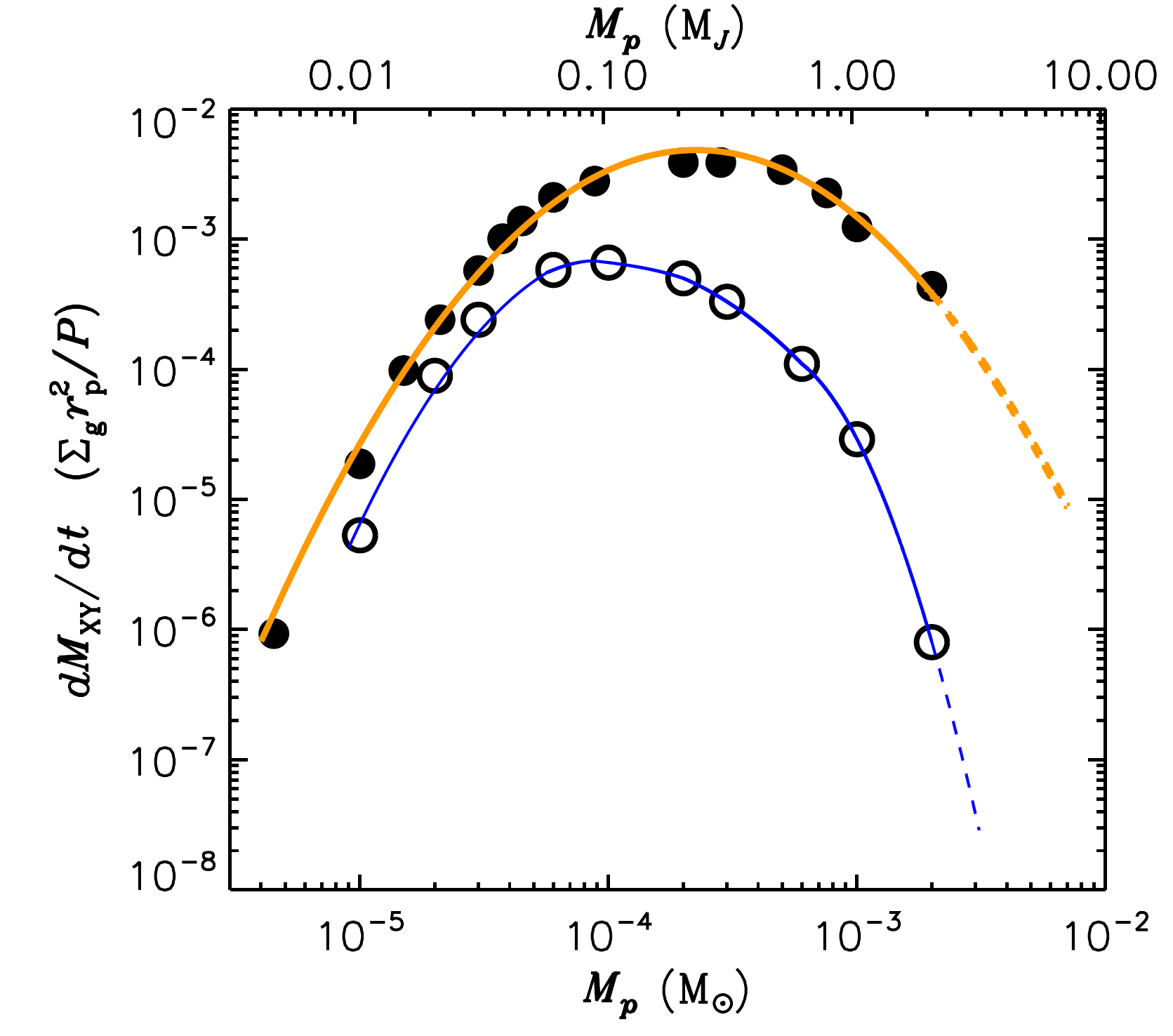}}
 \caption{Limiting accretion rates as a function of the planet's
          mass obtained from 3-D hydrodynamical calculations of
          a planet interacting with a circumstellar disk (see
          text for further details).
          Accretion rates are in units of the unperturbed surface
          density, $\Sigma_{\mathrm{g}}$, at the planet's orbital
          radius, $r_p$, and the planet's orbital period, $P$.
          Filled circles correspond to results for a disk with
          a viscosity $\alpha=4\times 10^{-3}$ at the planet's
          location. Empty circles are for a disk with
          $\alpha=4\times 10^{-4}$. The disk temperature at the
          orbital radius of the planet is $T\sim 100\;\mathrm{K}$.
          Masses on the lower axis are in units of the solar mass
          but they can be rescaled by the mass of the star,
          which is the unit of mass in the 3-D calculations.
          The thick orange curve represents a fit to the high 
          viscosity data according to Eq.~(\ref{eq:dotMxy}) 
          and the appropriate coefficients. The thin blue curve
          is a piecewise parabolic fit to the low viscosity data.
          }
          \label{fig:Mdot_xy}
\end{figure*}

Although we were able to halt the accretion of the planet at 
Jupiter's mass in a disk with $\alpha = 4 \times 10^{-3}$, 
this required either a very narrow gas feeding zone or a very 
special timing of disk dissipation (Section~\ref{sec:results}).
Thus, in order to find a more plausible mechanism for forming 
Jupiter-mass planets, we modeled planetary growth within a 
lower viscosity protoplanetary disk.
We performed simulations analogous to those described above 
for a planet within a disk of the same temperature but lower viscosity,  
$\alpha=4\times 10^{-4}$.  
The results for both viscosities are plotted in 
Fig.~\ref{fig:Mdot_xy}, together with the fit given by 
Eq.~(\ref{eq:dotMxy}) for $\alpha=4\times 10^{-3}$ and 
a piecewise parabolic fit (not given  here 
because it cannot be written in compact form) 
to accretion rates obtained from 
calculations with $\alpha=4\times 10^{-4}$.
In both cases,
accretion rates account for surface density perturbations which
depend on planet mass and disk viscosity.
Note that planets in the lower viscosity disk cannot accrete 
gas as rapidly as planets of the same mass within a disk that 
is ten times as viscous. This is because density perturbations 
are stronger in the lower viscosity case, even at small planet
masses. The difference in accretion rates is most profound for 
planets of Jupiter's mass and larger.
The shift of the peak accretion rates towards smaller planet
masses as kinematic viscosity decreases is in qualitative 
agreement with the results of \citet{tanigawa2007}.

So long as the gas density close to the planet's orbit remains
nearly undepleted, accretion rates in Fig.~\ref{fig:Mdot_xy}
can be understood in terms of gas accretion within the Bondi sphere,
at a rate $\dot{M}_{XY}\propto M^3_p$ 
for $M_p/M_{\star} \lesssim (H_p/r_p)^{3}/\sqrt{3}$, and accretion
within the Hill sphere at a rate $\dot{M}_{XY}\propto M_p$
for larger planet mass \citep[see][for details]{gennaro2008}. 
When density perturbations can no longer be neglected 
($R_{\mathrm{H}} \sim H_p$) and a gap starts to form, 
the accretion rate drops as the planet's mass increases.
The limiting gas accretion rates discussed here are not much affected 
by planet's migration as long as the migration timescale is larger than
the gap formation timescale.

\subsection{Effects of Circumsolar Disk 
Hydrodynamics on Jupiter's Accretion of Gas}
\label{sec:R_A}
We have modified our basic 1-D planet growth code to account 
for the limits of planet size and the supply of gas to the 
planet found with 3-D hydrodynamic calculations. 
As in \citetalias{bodenheimer2000b} and \citetalias{hubickyj2005}, 
we define the accretion radius as
\begin{equation}
R_A = \frac{G\, M_p}%
{c_s^2/k_{1} + G\,M_p/\left(k_{2}\,R_{\mathrm{H}}\right)},
\label{eq:RA}
\end{equation}
where $c_s$ is the sound speed in the disk
and $k_{1}$ and $k_{2}$ are constants.
In the limit of large $R_{\mathrm{H}}$, $R_A$ reduces to $k_{1}$
times the Bondi accretion radius. In the limit of small 
$R_{\mathrm{H}}$, $R_A$ reduces to $k_{2}\,R_{\mathrm{H}}$.
In previous studies, both $k_{1}$ and $k_{2}$ had been set to $1$.
In most of the simulations presented herein, we set $k_{1}=1$ and 
$k_{2}=1/4$, based on the calculations described in 
Section~\ref{sec:envsize}.
When thermal factors limit the gas accretion rate,
$\dot{M}_{XY}$ is obtained through the requirement that the 
computed radius of the planet, $R_p$, actually matches $R_A$.
When the hydrodynamics of the disk limits $\dot{M}_{XY}$, then
the radius $R_p$ is determined by the procedure outlined in
\citetalias{bodenheimer2000b}, and $R_p < R_A$.

In past work, the gas rate was arbitrarily capped at 
$1.053\times10^{-2}\,\mathrm{M}_{\oplus}/$yr to account
for the Bondi accretion limit.
In most of the present calculations, the hydrodynamic upper 
bound on the gas accretion rate is determined from the results 
shown in  Fig.~\ref{fig:Mdot_xy}.

\section{Parameters of our Simulations}
\label{sec:simupar}
In analogy with the principal simulation of 
\citetalias{hubickyj2005}, 
denoted 10\Linf\footnote{%
The run designator 10\Linf\ is used by \citetalias{hubickyj2005} 
to denote that 
the surface density of solids is $10\,\mathrm{g\,cm}^{-2}$, the 
dust opacity in the planet's envelope is 2\% that of the interstellar 
medium and there is no cutoff in accretion of planetesimals by 
the planet.  As these three properties hold for all of the 
simulations presented herein, we use a new set of designators, 
discussed below, for all of our new runs.}, 
all of our planetary evolution 
simulations are performed at $5.2\,\mathrm{AU}$ from a 
$1\,\mathrm{M}_{\odot}$ star within a disk that has an initial 
surface mass density of solids of 
$10\,\mathrm{g\,cm}^{-2}$, 
which are in the form of $100\,\mathrm{km}$ radius 
planetesimals\footnote{As pointed out by \citet{fortier2007}, 
using a more physically sophisticated model to compute the 
eccentricities and inclinations of planetesimals 
\citep{kokubo1998,kokubo2000,thommes2003} gives a lower value 
of $\dot{M}_Z$ for a given planetesimal size. However, growth 
of the core is more rapid with smaller planetesimals 
(\citetalias{pollack1996}; \citealp{fortier2007}). 
Thus, our simulations probably correspond more closely to a 
disk of $\sim 1$--$20\,\mathrm{km}$ radius planetesimals rather 
than the nominal size of $100\,\mathrm{km}$. 
The sizes of the planetesimals in all of these models have been 
chosen in a largely \textit{ad hoc} manner. A better estimate of 
$\dot{M}_Z$ requires more sophisticated calculations in which 
planetesimal sizes and velocities are computed in a 
self-consistent manner rather than merely prescribed.}, 
and of gas 
$\Sigma_{\mathrm{g}}(t=0) = 700\,\mathrm{g\,cm}^{-2}$.  
Also as in run 10\Linf, we assume that the opacity due to 
grains within the growing planet's envelope is $2$\% that of 
interstellar gas.  
In \citetalias{hubickyj2005}, 
we performed additional runs in order to examine 
the effects of differing values of the opacity of the 
planet's atmosphere and the initial surface mass density of 
solids in the disk, and of terminating the accretion of 
solids prior to the termination of gas accretion.  
Herein, we study the orthogonal processes of (1) varying 
our prescription for the physical size of the planet's 
envelope, (2) changing the formula for the maximum rate of 
gas accretion as a function of the planet's mass, and (3) 
reducing the gas surface  density of the protoplanetary
disk as a function of time throughout the accretion epoch.

The sixteen new simulations (runs) that we report herein
are divided into five groups. The input parameters for
each of these runs are listed in Table~\ref{table:1}.
\begin{table*}
 \caption{Input Parameters}\label{table:1}
 \centering
 \vspace*{1ex}%
 \resizebox{0.90\textwidth}{!}{%
 \small
 \begin{tabular}{|l||cccccc|}
 \hline
 Run    
 & $R_A$                          & $\sigma_{XY}$  (g/cm$^2$)  
 & $T_{\mathrm{neb}}$ (K)  & $\alpha$  
 & $\dot{M}_{XY,\mathrm{limit}}$ (M$_{\oplus}/$yr)   
 & limiting mass  \\
 \hline\hline
 10\Linf 
 &         escape to $R\sbs{H}$   & 700                       
 &  150                    & ---          
 &  $1.053\times 10^{-2}$    
 &  1 M$_{J}$     \\
 1G      
 &         escape to $R\sbs{H}$   & 700    
 &  150                    & $4\times 10^{-3}$
 &  Eq.~(\ref{eq:dotMxy})   
 &  1 M$_{J}$    \\
 \hline
 1s      
 &     escape to $0.25\,R\sbs{H}$ & 700    
 &  150                    & ---          
 &  $1.053 \times 10^{-2}$     
 &  1 M$_{J}$    \\
 1sG         
 &     escape to $0.25\,R\sbs{H}$ & 700    
 &  150                    & $4\times 10^{-3}$          
 &  Eq.~(\ref{eq:dotMxy})
 &  1 M$_{J}$    \\
 \hline
 1xsG        
 &  $0.25 \times$(escape to $R\sbs{H}$) & 700    
 &  150                    & $4\times 10^{-3}$
 &  Eq.~(\ref{eq:dotMxy})
 &  1 M$_{J}$    \\
 \hline
 2\textit{h} 
 &     escape to $0.25\,R\sbs{H}$ & 700       
 &  115                    & $4\times 10^{-3}$   
 &  Eq.~(\ref{eq:dotMxy})
 &  none           \\
 2\textit{h}J   
 &     escape to $0.25\,R\sbs{H}$ & 700       
 &  115                    & $4\times 10^{-3}$   
 &  Eq.~(\ref{eq:dotMxy})
 &  1 M$_{J}$   \\
 2\textit{l}    
 &     escape to $0.25\,R\sbs{H}$ & 700       
 &  115                    & $4\times 10^{-4}$   
 &  Fig.~\ref{fig:Mdot_xy}     
 &  none        \\
 2\textit{l}J   
 &     escape to $0.25\,R\sbs{H}$ & 700       
 &  115                    & $4\times 10^{-4}$   
 &  Fig.~\ref{fig:Mdot_xy}
 &  1 M$_{J}$   \\
 \hline
 3\textit{h}          
 &     escape to $0.25\,R\sbs{H}$ & Eq.~(\ref{eq:sig})
 &  115                    & $4\times 10^{-3}$    
 &  Eq.~(\ref{eq:dotMxy})
 &  none        \\
 3\textit{h}J         
 &     escape to $0.25\,R\sbs{H}$ & Eq.~(\ref{eq:sig})    
 &  115                    & $4\times 10^{-3}$    
 &  Eq.~(\ref{eq:dotMxy}) 
 &  1 M$_{J}$   \\
 3\textit{h}4R\sbs{H}
 &     escape to $0.25\,R\sbs{H}$ & Eq.~(\ref{eq:sig}) \& 4$R\sbs{H}$          
 &  115                    & $4\times 10^{-3}$    
 &  Eq.~(\ref{eq:dotMxy}) 
 &  none        \\
 3\textit{h}R\sbs{H}J   
 &     escape to $0.25\,R\sbs{H}$ & Eq.~(\ref{eq:sig}) \& 3.2$R\sbs{H}$  
 &  115                    & $4\times 10^{-3}$    
 &  Eq.~(\ref{eq:dotMxy})    
 &  1 M$_{J}$  \\
 3\textit{l}               
 &     escape to $0.25\,R\sbs{H}$ & Eq.~(\ref{eq:sig})
 &  115                    & $4\times 10^{-4}$   
 &  Fig.~\ref{fig:Mdot_xy}    
 &  none       \\
 3\textit{l}J     
 &     escape to $0.25\,R\sbs{H}$ & Eq.~(\ref{eq:sig})
 &  115                    & $4\times 10^{-4}$   
 &  Fig.~\ref{fig:Mdot_xy}
 &  1 M$_{J}$ \\
 3\textit{l}4R\sbs{H}         
 &     escape to $0.25\,R\sbs{H}$ & Eq.~(\ref{eq:sig}) \& 4$R\sbs{H}$     
 &  115                    & $4\times 10^{-4}$   
 &  Fig.~\ref{fig:Mdot_xy}
 &  none       \\
 3\textit{l}R\sbs{H}J        
 &     escape to $0.25\,R\sbs{H}$ & Eq.~(\ref{eq:sig}) \& 11R\sbs{H}   
 &  115                    & $4\times 10^{-4}$   
 &  Fig.~\ref{fig:Mdot_xy}
 &  1 M$_{J}$  \\
 \hline
 \end{tabular}
                              }
\end{table*}

In the three groups of runs whose labels begin with ``1'', 
referred to collectively as ``groups 1'',
$T=150\,\mathrm{K}$ and the gas surface density remains constant 
throughout the computation, as in \citetalias{hubickyj2005}.
The purpose of these simulations is to determine the effects 
of restricting gas flow onto the planet using the formula 
given by 3-D accretion simulations and/or reducing the radius 
of the outer boundary of the planet's envelope by up to a 
factor of four to account approximately for unbound  
protoplanetary disk gas flowing through the planet's Hill 
sphere (Fig.~\ref{fig:tracers_xy}). 
Specifically, Run~1G begins with the evolution found in 
10\Linf, but uses the gas flow limits from the 3-D 
calculations (Eq.~\ref{eq:dotMxy}) in place of the 
constant maximum gas accretion rate 
$\dot{M}_{XY}=1.053\times10^{-2}\,\mathrm{M}_{\oplus}$ 
per year that 
was inspired by a Bondi-type estimate and used in our 
previous studies. Run~1s uses the same formulation as 
10\Linf\ except that the 
planet's accretion radius, $R_A$, (which prior to the onset
of hydrodynamic limits to $\dot{M}_{XY}$ lies at the
outer boundary of the planet's envelope) 
is placed at the radius where the gas has enough
thermal energy to escape to $R_{\mathrm{H}}/4$ rather than 
to $R_{\mathrm{H}}$. Run~1sG has the same planet size as 
Run~1s and the same limits on gas flow as in simulation 1G.
Run~1xsG provides a test of the sensitivity of moving the 
outer boundary of the envelope significantly inwards 
throughout the planet's evolution.
In Run~1xsG we (arbitrarily) set the outer boundary of the 
envelope to be $1/4$ as far from the center of the planet 
as in 10\Linf\ and use the same limits on the gas flow 
as in Runs~1G and 1sG 
(i.e., we set $k_1=k_2=1/4$ in Eq.~\ref{eq:RA}).
In order to produce planets with mass equal to that of 
Jupiter, we taper off the accretion rate by multiplying the 
calculated maximum rate of gas accretion by a function which 
begins at unity when $M_p = 0.85\,\mathrm{M}_J$ and drops 
linearly with the planet's mass so that it vanishes when 
$M_p = 1\,\mathrm{M}_J$; this is the same procedure as was 
used by \citetalias{hubickyj2005}.

All simulations in groups 2 and 3 assume the temperature of 
the disk (and thus at the upper boundary of the planet as 
long as the planet's envelope remains in contact with the 
disk) is $115\,\mathrm{K}$, the value used in the 3-D 
calculations for a gas of mean molecular weight $2.25$.
The planet radius prescription for all of these groups is 
the same as used for Runs~1s and 1sG.  
 
In group 2, as in groups 1, the surface density of the gas 
remains constant at 
$\Sigma_{\mathrm{g}} = 700\,\mathrm{g\,cm}^{-2}$. 
In group 3, the surface density of gas within the disk 
drops linearly according to the formula: 
\begin{equation}
\Sigma_{\mathrm{g}}(t)=\left\{%
\begin{array}{ccc}
   700\,\mathrm{g\,cm}^{-2}\,\left(%
   \frac{3\,\mathrm{Myr}- t}{3\,\mathrm{Myr}}\right) & \mathrm{if} & t\le 3\;\mathrm{Myr}\\
             0                                                                        & \mathrm{if} & t > 3\;\mathrm{Myr}
\end{array}
                                           \right.\,.
\label{eq:sig}
\end{equation}
The above equation accounts for disk dispersal in the planet formation
region over timescales suggested by observations \citep{haisch2001} and
recent theoretical models of disk photoevaporation due to combined
action of FUV, EUV, and X-ray photons emitted by a solar-mass 
star \citep{gorti2008}.

In the ``\textit{h}'' runs within groups 2 and 3, the 
dimensionless disk viscosity is assumed to be 
$\alpha=4\times 10^{-3}$, and therefore the limiting gas 
accretion rate is taken from the 3-D calculations for disks 
with this viscosity (as in Runs~1G, 1sG and 1xsG), as given 
by Eq.~(\ref{eq:dotMxy}). 
For the ``\textit{l}'' runs within groups 2 and 3, the lower 
viscosity disk, $\alpha=4\times 10^{-4}$, is assumed, and the 
piecewise parabolic fits shown as a narrow blue curve in 
Fig.~\ref{fig:Mdot_xy} are used for the upper bound of the 
planet's gas accretion rate.

In Runs~3\textit{h} and 3\textit{l}, the planet continues to 
accrete for the entire $3\,\mathrm{Myr}$ that gas is assumed 
to be present.
Analogously, the planet's accretion is halted (in this case 
abruptly) after $3\,\mathrm{Myr}$ in Runs~2\textit{h} and
2\textit{l}. 
Accretion is terminated when the planet reaches 
$1\,\mathrm{M}_J$ in Runs~2\textit{h}J, 2\textit{l}J, 3\textit{h}J 
and 3\textit{l}J.
For Runs~2\textit{h}J, 2\textit{l}J, and 3\textit{h}J, 
the same taper as used for groups 1 was applied, but using 
this prescription for Run~3\textit{l}J did not provide adequate 
mass to the planet, so we began the linear tapering when 
the planet mass was $0.975\,\mathrm{M}_J$ rather than 
$0.85\,\mathrm{M}_J$.

We account for depletion of the gas in the disk via accretion 
onto the planet as well as the linear decline from the overall 
dissipation of the protoplanetary disk in 
Runs~3\textit{h}4R$_\mathrm{H}$, 3\textit{h}R$_\mathrm{H}$J,
3\textit{l}4R$_\mathrm{H}$ and 3\textit{l}R$_\mathrm{H}$J. 
However, gas pressure gradients can act
to replace accreted gas, so rather than taking the gas loss rate 
within the planet's gas feeding zone to be the sum of 
(\textit{i}) the planet's accretion of gas and 
(\textit{ii}) the linear gas surface density drop-off 
rate assumed for the 
overall disk multiplied by the area of the planet's gas feeding 
zone, we take the instantaneous gas loss rate to be the larger 
of these two quantities.
We take the size of the planet's gas feeding zone to be 
proportional to the size of its Hill sphere, so the gas feeding 
zone expands with the growth of the planet (Eq.~\ref{eq:RH}) 
into regions not depleted by previous accretion by the
planet. Note that these more distant regions are still affected 
by overall gas removal of the disk according to Eq.~(\ref{eq:sig}). 
We account for the expansion of the planet's gas feeding zone 
similarly to our formula for computing the surface density of 
solids \citepalias[Section~2.1 of][]{pollack1996}.   
In Runs~3\textit{h}4R\sbs{H} and 3\textit{l}4R\sbs{H}, 
the half-width of the planet's gas accretion zone is taken to be 
$4\,R_\mathrm{H}$, the same as its solids accretion zone, and gas 
is allowed to accrete until there is no gas left.  
In Runs~3\textit{h}R$_\mathrm{H}$J and 3\textit{l}R\sbs{H}J, 
we choose (by an iterative procedure) the size of the gas accretion 
zone to be $3.2\,R_\mathrm{H}$ and $11\,R_\mathrm{H}$, 
respectively, so that the feeding zone runs out of gas when the 
planet's mass is $\approx 1\,\mathrm{M}_J$. 

\section{Results}
\label{sec:results}
In this section, we present the results of all of the simulations 
described in Section~\ref{sec:simupar}. Readers interested in a 
summary of these results may skip to Section~\ref{sec:summary}, 
which includes a figure displaying the temporal evolution of the 
mass, radius and luminosity of the planet in what is probably our 
most physically realistic model of the growth of Jupiter, 
Run~3\textit{l}R\sbs{H}J.

\begin{table*}[t!]
 \caption{Results: Phases~I and II}\label{table:2}
 \centering
 \vspace*{1ex}%
 \resizebox{0.90\textwidth}{!}{%
 \small
 \begin{tabular}{|l|l||c|c|c|c|c|}
 \cline{1-1}
 \cline{2-2}
 \cline{3-3}
 \cline{4-4}
 \cline{5-5}
 \cline{6-6}
 \cline{7-7}
 \hline
 \cline{1-1}
 \cline{2-2}
 \cline{3-3}
 \cline{4-4}
 \cline{5-5}
 \cline{6-6}
 \cline{7-7}
 \hline
 &                                & 10\Linf/1G           & 1s/1sG    
 & 1xsG                  & 2      & 3      \\
 \hline\hline
 \cline{1-1}
 \cline{2-2}
 \cline{3-3}
 \cline{4-4}
 \cline{5-5}
 \cline{6-6}
 \cline{7-7}
 \hline
 \raisebox{-8.5ex}[+0.0ex][-0.0ex]{\parbox{5.5em}{%
 \textsc{First\\ Luminosity\\ Peak}}}
 & Time\sps{a}                    & 0.348                & 0.348
 &  0.352                & 0.345  & 0.349  \\
 & $M_{Z}$\sps{b}                 & 7.98                 & 8.04
 &  8.15                 & 8.06   & 8.35   \\
 & $M_{XY}$\sps{b}                & 0.014                & 0.011 
 &  0.0093               & 0.013  & 0.014  \\
 & $\dot{M}_Z$                    & $8.8 \times 10^{-5}$ & $8.8 \times 10^{-5}$
 & $8.5 \times 10^{-5}$  & $8.8 \times 10^{-5}$ & $8.5\times 10^{-5}$ \\
 & $\log{L}$\sps{d}               & -5.05                & -5.05
 & -5.06                 & -5.05  & -5.05  \\
 & $R_p$\sps{e}                   & 41.53                & 25.56
 & 10.51                 & 29.08  & 29.62  \\
 \hline
 \raisebox{-8.5ex}[+0.0ex][-0.0ex]{\parbox{5.5em}{%
 \textsc{End of\\ Phase~I}}}
 & Time\sps{a}                    & 0.429                & 0.429
 & 0.436                 & 0.427  & 0.426  \\
 & $M_{Z}$\sps{b}                 & 11.49                & 11.47
 & 11.47                 & 11.49  & 11.48  \\
 & $M_{XY}$\sps{b}                & 0.32                 & 0.29
 & 0.27                  & 0.32   & 0.30   \\ 
 & $\dot{M}_Z$\sps{c}             & $8.9\times 10^{-6}$  & $8.7\times 10^{-6}$
 & $7.7\times 10^{-6}$   & $8.5 \times 10^{-6}$ & $8.8\times 10^{-6}$ \\
 & $\log{L}$\sps{d}               & -5.90                & -5.91
 & -5.95                 & -5.92  & -5.90  \\
 & $R_p$\sps{e}                   & 56.78                & 31.34
 &  14.23                & 34.90  & 36.24  \\
 \hline
 \raisebox{-8.5ex}[+0.0ex][-0.0ex]{\parbox{5.5em}{%
 \textsc{Mid\\ Phase~II}}}
 & Time\sps{a}                    & 1.59                 & 1.69
 & 1.86                 & 1.61    & 1.63   \\ 
 & $M_{Z}$\sps{b}                 & 14.0                 & 14.0
 & 14.0                 & 14.0    & 14.0   \\ 
 & $M_{XY}$\sps{b}                & 7.0                  & 7.0
 & 7.0                  & 7.0     & 7.0    \\
 & $\dot{M}_Z$                    & $2.2\times 10^{-6}$  & $2.0 \times 10^{-6}$
 & $1.8 \times 10^{-6}$ & $2.2\times 10^{-6}$ & $2.1\times 10^{-6}$  \\
 & $\dot{M}_{XY}$\sps{c}          & $7.8\times 10^{-6}$  & $7.3\times 10^{-6}$
 & $6.5\times 10^{-6}$  & $7.6\times 10^{-6}$ & $7.4\times 10^{-6}$  \\
 & $\log{L}$\sps{d}               & -6.41                & -6.45
 & -6.49                & -6.44   & -6.44  \\
 & $R_p$\sps{e}                   & 90.92                & 45.23 
 & 23.16                & 49.37   & 48.49  \\
 \hline
 \raisebox{-8.5ex}[+0.0ex][-0.0ex]{\parbox{5.5em}{%
 \textsc{Crossover\\ Point}}}
 & Time\sps{a}                    & 2.28                 &  2.38
 & 2.59                 & 2.30    & 2.32   \\
 & $M_{\mathrm{cross}}$\sps{b}    & 16.15                & 16.16
 & 16.16                & 16.16   & 16.16  \\
 & $\dot{M}_Z$\sps{c}             & $5.0\times 10^{-6}$  & $5.4\times 10^{-6}$ 
 & $4.7\times 10^{-6}$            & $5.4\times 10^{-6}$  & $5.4\times 10^{-6}$
                                           \\
 & $\dot{M}_{XY}$\sps{c}
 & $2.7\times 10^{-5}$            & $2.7 \times 10^{-5}$ & $2.3 \times 10^{-5}$
 & $2.7\times 10^{-5}$            & $2.7 \times 10^{-5}$
                                           \\
 & $\log{L}$\sps{d} & -6.07                              & -6.08
 & -6.09                & -6.07   & -6.07  \\
 & $R_p$\sps{e}                   & 125.8                & 56.60
 & 31.30                & 60.03   & 59.72  \\
 \hline
 \raisebox{-0.0ex}[+0.0ex][-0.0ex]{\parbox{5.5em}{%
 \resizebox{5.4em}{!}{%
 \textsc{Bifurcation}}}}
 & Time\sps{a}                    & 2.367                &  2.488
 & N/A                  & 2.446   & 2.437  \\
 \hline
 \end{tabular}
                              }\\[1ex]
 \hspace{0.2in}%
 \begin{minipage}[t]{\linewidth}
 \footnotesize
 \noindent%
 \sps{a} Time is in units of millions of years, Myr.\\
 \sps{b} Mass is in units of Earth's mass, M$_{\oplus}$.\\
 \sps{c} The accretion rate is in units of Earth masses per 
         year, M$_{\oplus}/$yr.\\
 \sps{d} Luminosity is in units of solar luminosity, 
         L$_{\odot}$.\\
 \sps{e} Radius is in units of Jupiter's present equatorial radius, R$_J$.
 \end{minipage}
\end{table*}
\begin{table*}[th!]
 \caption{Results: Phase~III, Groups 1 \& 2}\label{table:3}
 \centering
 \vspace*{1ex}%
 \resizebox{0.90\textwidth}{!}{%
 \small
 \begin{tabular}{|l|l||ll|ll|l||llll|}
 \cline{3-11}
 \hline
 \cline{3-11}
 \hline  
 &                                
 & 10\Linf  & 1G     & 1s       & 1sG     & 1xsG  
 & 2\textit{h}       & 2\textit{h}J      
 & 2\textit{l}       & 2\textit{l}J                           \\
 \hline
 \cline{3-11}
 \hline\hline
 \raisebox{-6.5ex}[+0.0ex][-0.0ex]{\parbox{5.5em}{%
 \textsc{Onset\\ of Limited\\ Gas\\ Accretion}}}
 & Time\sps{a}                    
 & 2.367    & 2.367  & 2.488    & 2.488   & 2.710  
 &    \multicolumn{2}{c}{2.446} &  \multicolumn{2}{c|}{2.446} \\
 & $M_{Z}$\sps{b}        
 & 16.62    & 16.62  & 16.79    & 16.79   & 16.89
 &    \multicolumn{2}{c}{17.03} &  \multicolumn{2}{c|}{17.03} \\
 & $M_{XY}$\sps{b}       
 & 44.33    & 59.18  & 48.43    & 58.80   & 60.12   
 &    \multicolumn{2}{c}{64.18} &  \multicolumn{2}{c|}{55.47} \\
 & $\dot{M}_{XY}$\sps{c} 
 & 0.0105   & 0.277  & 0.0105   & 0.277   & 0.277 
 &    \multicolumn{2}{c}{0.275} &  \multicolumn{2}{c|}{0.0278} \\
 & $R_p$\sps{e}
 & 182.7    & 163.2  & 76.0     & 52.3    & 34.3    
 &    \multicolumn{2}{c}{68.0}  &  \multicolumn{2}{c|}{72.8}  \\
 \hline
 \raisebox{-5.5ex}[+0.0ex][-0.0ex]{\parbox{5.5em}{%
 \textsc{Second\\ Luminosity\\ Peak}}}
 & Time\sps{a}
 &  2.368   & 2.368  & 2.507    & 2.488   & 2.710
 &  2.447   & 2.447             &  \multicolumn{2}{c|}{2.451} \\
 & $M_{XY}$\sps{b}
 & 254.3    & 215.2  & 254.2    & 144.8   & 180.0
 & 290.7    & 252.2             &  \multicolumn{2}{c|}{123.9} \\
 & $\log{L}$\sps{d}
 & -2.34    & -1.47  & -2.36    & -1.53   & -1.48 
 & -1.43    & -1.44             &  \multicolumn{2}{c|}{-2.78} \\
 & $R_p$\sps{e}
 & 1.73     & 2.48   & 1.80     & 2.93    & 2.61 
 & 2.02     & 2.17              &  \multicolumn{2}{c|}{2.95}  \\
 \hline
 \raisebox{-5.5ex}[+0.0ex][-0.0ex]{\parbox{5.5em}{%
 \textsc{Accretion\\ Stops}}}
 & Time\sps{a}
 &  2.421   & 2.374  & 2.541    & 2.494   & 2.716
 & 3.00     &  2.45             & 3.00    & 2.60              \\
 & $M_{Z}$\sps{b}
 & 16.89    & 16.65  & 17.10    & 16.83   & 16.93
 & 20.36    & 17.07             & 20.36   & 17.96             \\
 & $M_{XY}$\sps{b}
 & 301.8    & 302.0  & 301.5    & 301.8   & 301.7
 & 1810.0   & 301.6             & 525.0   & 300.0             \\
 & $R_p$\sps{e}
 & 1.63     & 1.76   & 1.68     & 1.74    & 1.68 
 & 1.34     & 1.74              & 1.57    & 1.78              \\
 \hline
 \end{tabular}
                              }\\[1ex]
 \hspace{0.2in}%
 \begin{minipage}[t]{\linewidth}
 \footnotesize
 \noindent%
 \sps{a} Time is in units of millions of years, Myr.\\
 \sps{b} Mass is in units of Earth's mass, M$_{\oplus}$.\\
 \sps{c} The accretion rate is in units of Earth masses per 
         year, M$_{\oplus}/$yr.\\
 \sps{d} Luminosity is in units of solar luminosity, 
         L$_{\odot}$.\\
 \sps{e} Radius is in units of Jupiter's present equatorial radius, R$_J$.
 \end{minipage}
\end{table*}
\begin{table*}[th!]
 \caption{Results: Phase~III, Groups 3}\label{table:4}
 \centering
 \vspace*{1ex}%
 \resizebox{0.90\textwidth}{!}{%
 \small
 \begin{tabular}{|l|l||llllllll|}
 \cline{3-10}
 \hline
 \cline{3-10}
 \hline
 &  
 & 3\textit{h}            & 3\textit{h}J         & 3\textit{h}4R\sbs{H}  
 & 3\textit{h}R\sbs{H}J   & 3\textit{l}          & 3\textit{l}J  
 & 3\textit{l}4R\sbs{H}   & 3\textit{l}R\sbs{H}J     \\
 \hline
 \cline{3-10}
 \hline\hline
 \raisebox{-6.5ex}[+0.0ex][-0.0ex]{\parbox{5.5em}{%
 \textsc{Onset\\ of Limited\\ Gas\\ Accretion}}}
 & Time\sps{a}
 & \multicolumn{2}{c}{2.455}                     & 2.454
 & 2.454                  & \multicolumn{2}{c}{2.453}
 & 2.453                  & 2.453                    \\
 & $M_{Z}$\sps{b}
 & \multicolumn{2}{c}{16.93}                     & 16.92
 & 16.92                  & \multicolumn{2}{c}{16.91} 
 & 16.91                  & 16.91                    \\
 & $M_{XY}$\sps{b}
 & \multicolumn{2}{c}{56.44}                     & 54.01
 & 51.36                  & \multicolumn{2}{c}{39.19}
 & 40.86                  & 40.87                   \\
 & $\dot{M}_{XY}$\sps{c}  
 & \multicolumn{2}{c}{0.0522}                    & 0.0471
 & 0.0459                 & \multicolumn{2}{c}{0.00596}
 & 0.00559                & 0.00582                 \\
 & $R_p$\sps{e}
 & \multicolumn{2}{c}{67.6}                      & 79.9  
 & 75.0                   & \multicolumn{2}{c}{75.9}
 & 75.8                   & 76.0                     \\
 \hline
 \raisebox{-5.5ex}[+0.0ex][-0.0ex]{\parbox{5.5em}{%
 \textsc{Second\\ Luminosity\\ Peak}}}
 & Time\sps{a}
 & 2.461                  & 2.461                & 2.458 
 & 2.457                  & \multicolumn{2}{c}{2.481}
 & 2.477                  & 2.480                    \\
 & $M_{XY}$\sps{b}
 & 260.7                  & 258.2                & 163.4
 & 169.4                  & \multicolumn{2}{c}{130.5}
 & 108.3                  & 125.9                    \\
 & $\log{L}$\sps{d}
 & -2.07                  & -2.07                & -2.41
 & -2.49                  & \multicolumn{2}{c}{-3.48}
 & -3.56                  & -3.51                    \\
 & $R_p$\sps{e}
 & 1.85                   & 1.85                 & 2.56 
 & 2.71                   & \multicolumn{2}{c}{2.39}
 & 2.99                   & 2.43                     \\
 \hline
 \raisebox{-5.5ex}[+0.0ex][-0.0ex]{\parbox{5.5em}{%
 \textsc{Accretion\\ Stops}}}
 & Time\sps{a}
 & 2.998                  & 2.476                & 2.549 
 & 2.517                  & 2.998                & 2.914
 & 2.825                  & 2.964                    \\
 & $M_{Z}$\sps{b}
 & 20.12                  & 17.05                & 17.49
 & 17.30                  & 20.18                & 19.68
 & 19.15                  & 19.98                    \\
 & $M_{XY}$\sps{b}
 & 1072.0                 & 301.6                & 385.6
 & 301.0                  & 309.1                & 297.8
 & 244.1                  & 297.6                    \\
 & $R_p$\sps{e}
 & 1.39                   & 1.67                 & 1.66 
 & 1.72                   & 1.62                 & 1.64
 & 1.80                   & 1.64                     \\
\hline
 \end{tabular}
                              }\\[1ex]
 \hspace{0.2in}%
 \begin{minipage}[t]{\linewidth}
 \footnotesize
 \noindent%
 \sps{a} Time is in units of millions of years, Myr.\\
 \sps{b} Mass is in units of Earth's mass, M$_{\oplus}$.\\
 \sps{c} The accretion rate is in units of Earth masses per 
         year, M$_{\oplus}/$yr.\\
 \sps{d} Luminosity is in units of solar luminosity, 
         L$_{\odot}$.\\
 \sps{e} Radius is in units of Jupiter's present equatorial radius, R$_J$.
 \end{minipage}
\end{table*}
Within the five individual groups, all of the runs use the same 
computations for 
the initial phases of planet growth. 
Properties of the planet at milestone times during the early
(thermally-regulated) phases of its growth (denoted Phase~I and 
Phase~II) for all of the 
simulations are presented in Table~\ref{table:2}. 
Note that for this epoch, only five simulations are required, and 
one of these is taken from the 10\Linf\ run of 
\citetalias{hubickyj2005}.
Table~\ref{table:2} also lists the ``bifurcation'' time, defined as the
time at which the runs began to differ due to the differing 
assumptions regarding supply of gas to the planet.
(The Run~1xsG has no bifurcation time because it was the only hypothesis
considered within its group.)
Analogous information for each of the runs at later times (at the
onset of hydrodynamic limits to accretion, at maximum luminosity,
and at the termination of accretion; Phase~III) is displayed in 
Tables~\ref{table:3} and 
\ref{table:4}.  

\begin{figure*}[t!]
 \centering
 \resizebox{0.65\hsize}{!}{\includegraphics[angle=90]{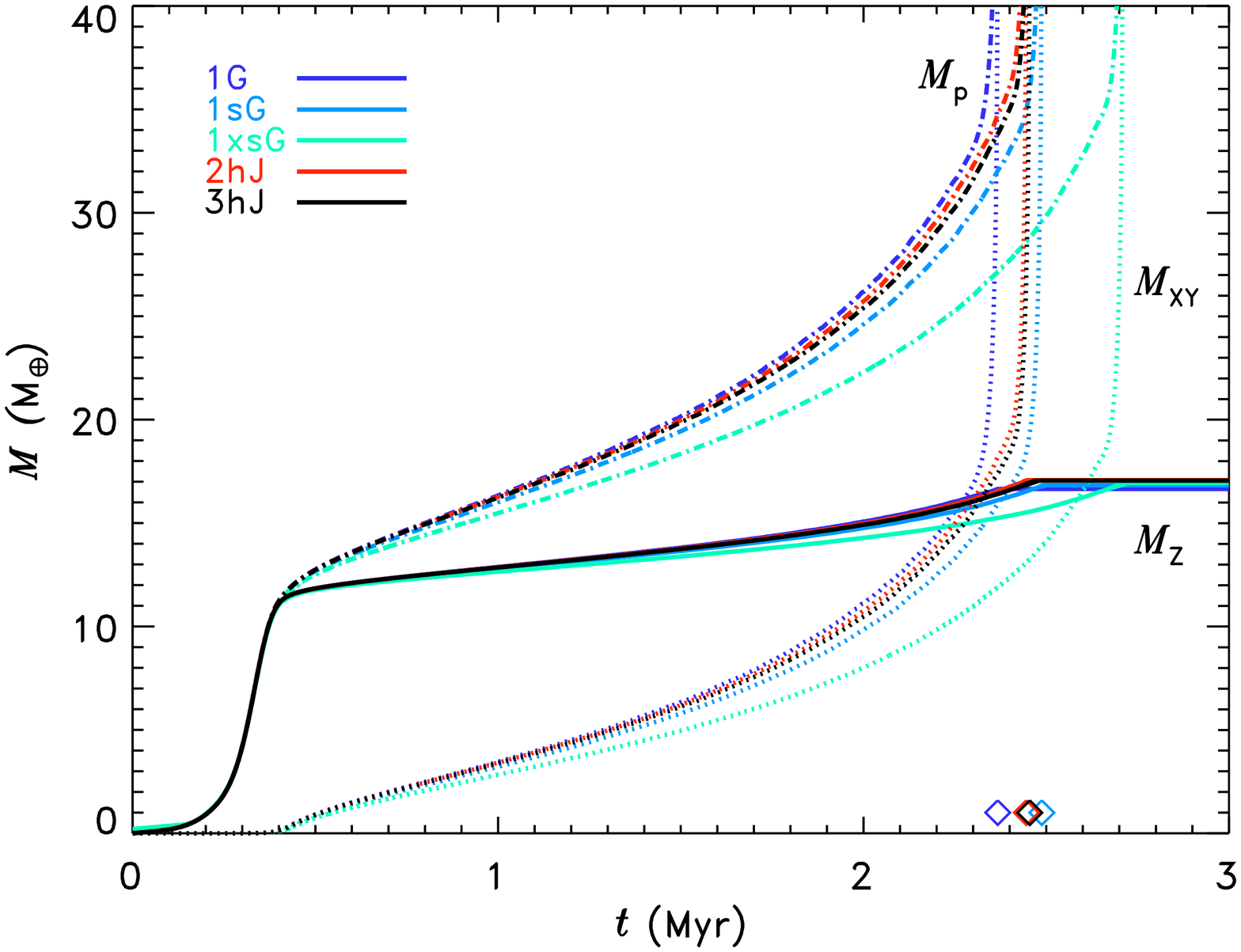}}
 \caption{Mass of the giant planet as a function of time for the
          five runs listed in the color key, all of which use 
          Eq.~(\ref{eq:dotMxy}) to specify the hydrodynamic cap
          of gas accretion rates. Solid lines denote the mass of the 
          condensible component of the planet, $M_Z$, dotted lines 
          the H/He component, $M_{XY}$, and dot dashed lines the 
          planet's total mass, $M_p$. The ordinate is truncated at 
          $40\,\mathrm{M}_{\oplus}$ in order to show details of the
          evolution prior to gas runaway. The bifurcation times for
          each of
          the four simulations with multiple endings are denoted by
          the diamonds situated just above the abscissa.
          Apart from small changes in the solid lines at later times
          and in the very last portions of some of the dotted lines,
          the results that are plotted here are applicable to all
          of the runs listed in Tables~\ref{table:3} and \ref{table:4}.
          }
          \label{fig:mpe}
\end{figure*}
\begin{figure*}[th!]
 \centering
 \resizebox{0.65\hsize}{!}{\includegraphics[angle=90]{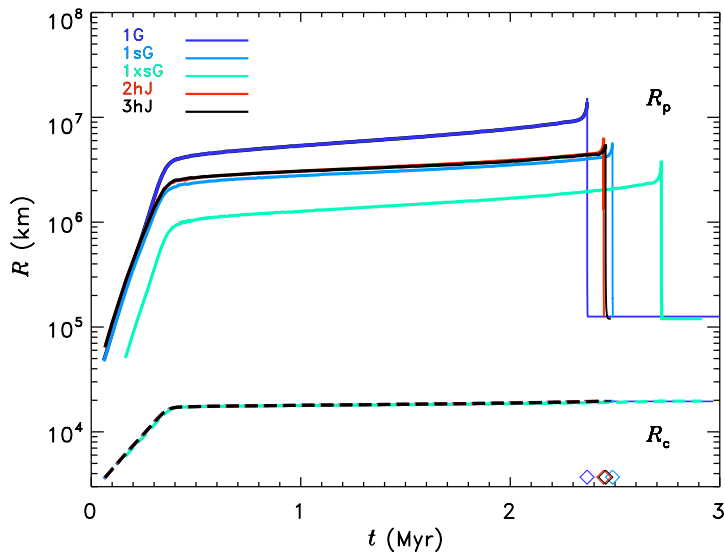}}
 \caption{\textit{Solid curves:} Outer radius of the planet as a 
          function of time for the same five runs whose masses are
          shown in Fig.~\ref{fig:mpe}. The thicker portions of four
          curves correspond to times prior to the bifurcation
          of the runs. The primary differences between the
          planet's radius from one run to another prior to the onset
          of limited gas accretion are direct consequences of the
          various prescriptions that were used for planetary radius 
          (see Table~\ref{table:1}). The slight differences in 
          $R_p$ among runs 1sG, 2\textit{h}J, and 3\textit{h}J
          result from differences in the growth rates of these
          planets (compare with Fig.~\ref{fig:mpe}).
          \textit{Dashed curves:} The core radius, $R_c$, as a 
          function of time for the same runs. 
          As in Fig.~\ref{fig:mpe}, diamonds denote bifurcation
          times.
          }
          \label{fig:rpe}
\end{figure*}
\begin{figure*}[th!]
 \centering
 \resizebox{0.65\hsize}{!}{\includegraphics[angle=90]{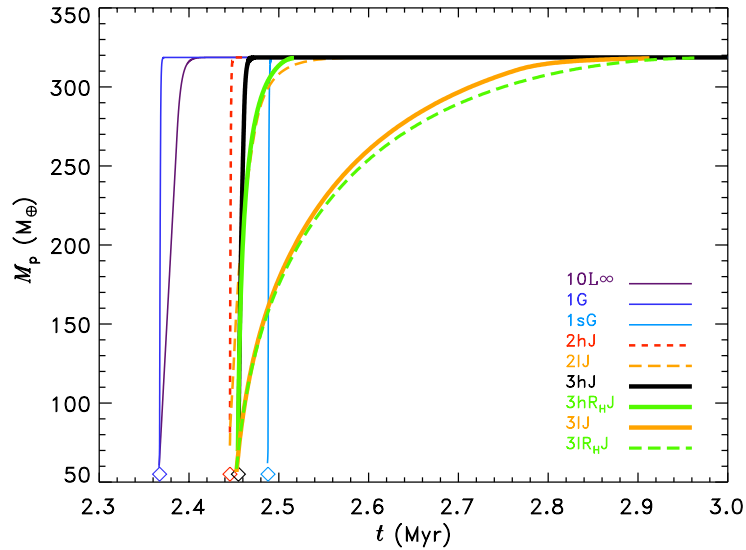}}
 \caption{Total mass of the planet is shown at late times for the runs
          specified. Diamonds denote bifurcation times.
          Note that the lower value of 
          the disk viscosity (Runs~2\textit{l}J, 3\textit{l}J, and
          3\textit{l}R\sbs{H}J)
          produces more gradual (and we believe more
          realistic) termination of accretion for $1\,\mathrm{M}_J$
          planets.
          }
          \label{fig:mpl}
\end{figure*}
\begin{figure*}[th!]
 \centering
 \resizebox{0.65\hsize}{!}{\includegraphics[angle=90]{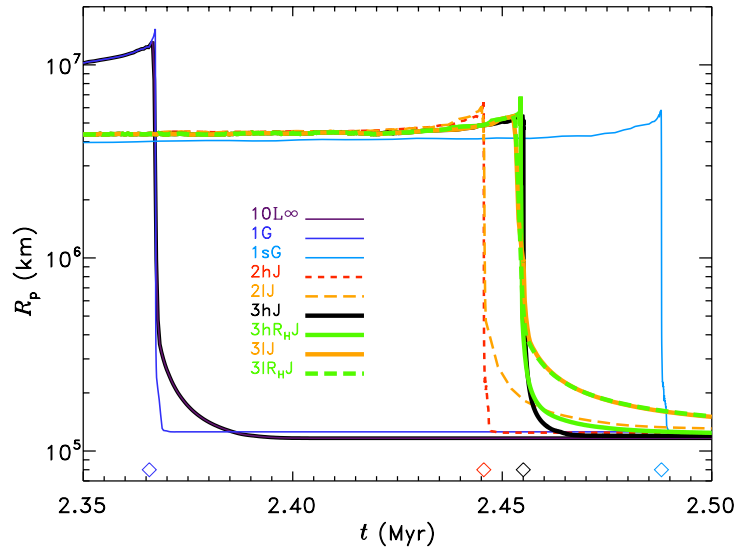}}
 \caption{The outer radius of the planet as a function of time at 
          late times, for the same runs whose late-time masses 
          are displayed in Fig.~\ref{fig:mpl}. The radii of these 
          planets change little during the interval 
          $2.5\,\mathrm{Myr}<t<3\,\mathrm{Myr}$ (not shown in the
          plot).
          Diamonds denote bifurcation times.
          }
          \label{fig:rpl}
\end{figure*}
\begin{figure*}[th!]
 \centering
\resizebox{0.65\hsize}{!}{\includegraphics[angle=90]{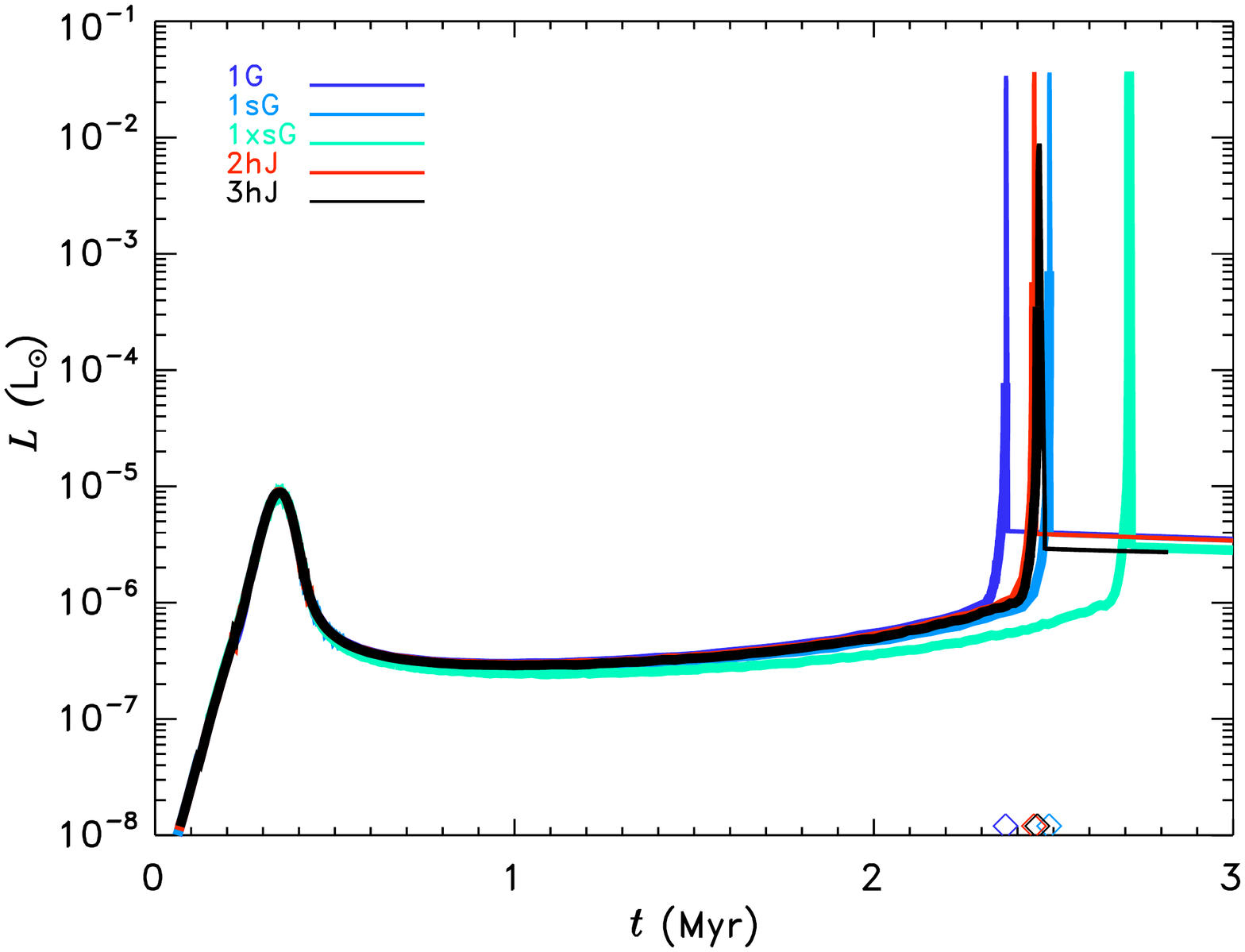}}
\resizebox{0.65\hsize}{!}{\includegraphics[angle=90]{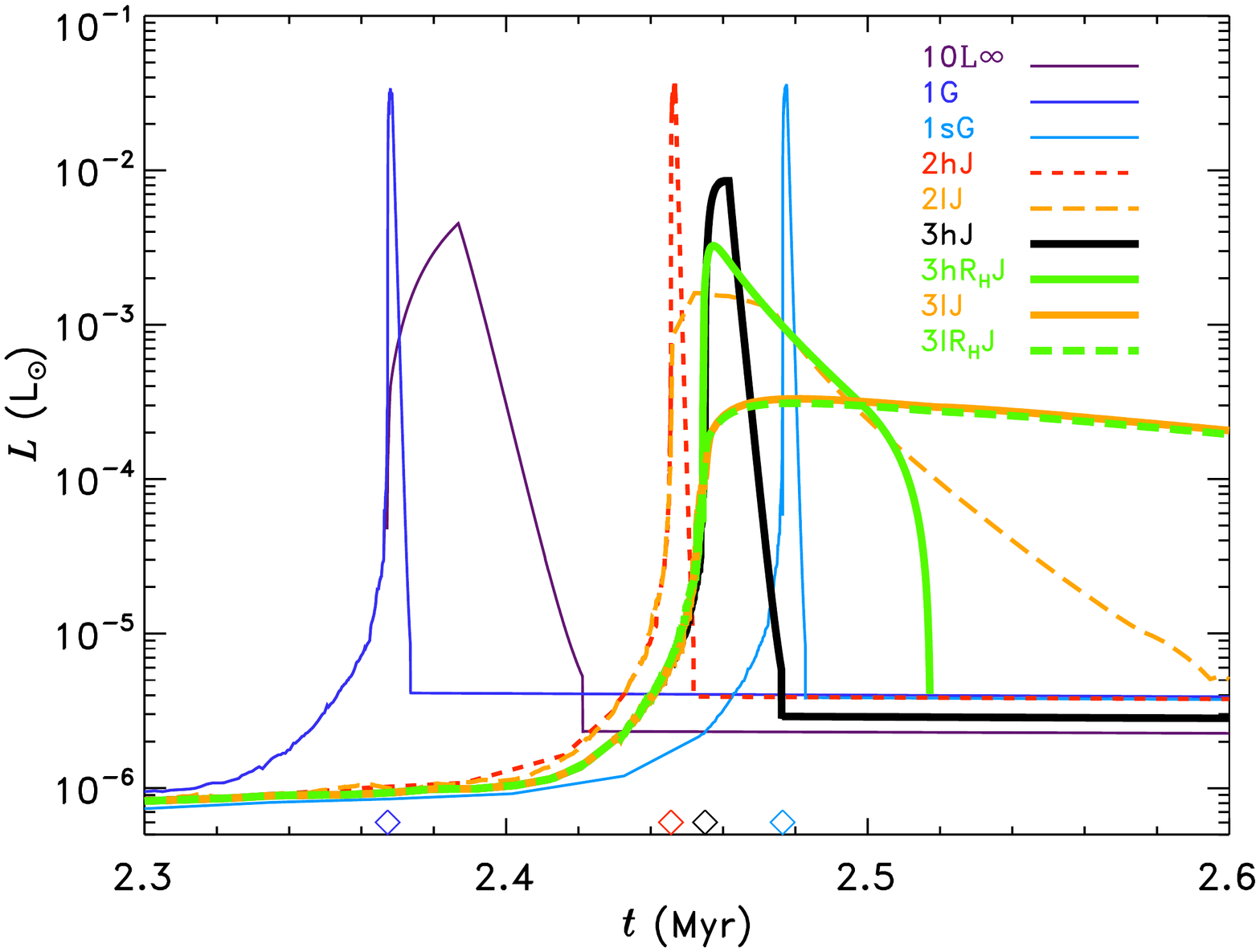}}
 \caption{The planet's luminosity as a function of time 
          is shown for all of the runs which resulted in planets 
          of mass equal to that of Jupiter. 
          Diamonds denote bifurcation times.
          Top: The companion
          to Figs.~\ref{fig:mpe} and \ref{fig:rpe}, shows data
          from five pre-bifurcation runs as thick solid lines and
          post-bifurcation results from selected cases in high 
          viscosity disks as narrow lines.
          Bottom: The companion to Figs.~\ref{fig:mpl} and 
          \ref{fig:rpl}, shows the post-bifurcation luminosity of
          nine runs that produce planets of mass $1\,\mathrm{M}_J$.
          Note in all cases the steep increase in 
          luminosity as the rate of gas accretion accelerates
          after crossover; this is a real physical consequence of
          the core nucleated accretion model of giant planet 
          formation. 
          The value at which luminosity peaks depends
          upon the planetary mass at which disk hydrodynamics begins
          to limit the rate of mass accretion, and thus on the 
          viscosity and surface density of the disk in the vicinity 
          of the planet. Those simulations in which accretion of gas 
          is tapered off exhibit a corresponding taper in luminosity;
          the curve for run 3\textit{l}R\sbs{H}J is probably most 
          realistic, as this run has the most plausible treatment of 
          the tail off in gas accretion.
          }
          \label{fig:lp}
\end{figure*}
\begin{figure*}[th!]
 \centering
\resizebox{0.65\hsize}{!}{\includegraphics[angle=90]{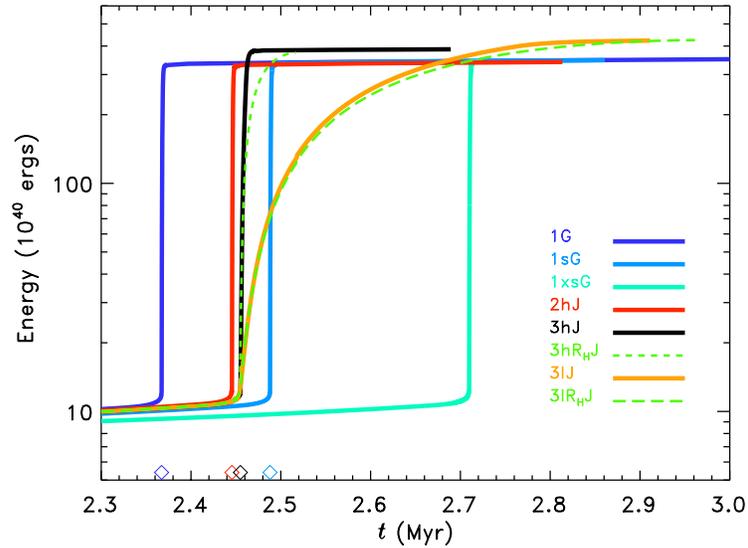}}
 \caption{Integrated luminosity (i.e., total energy radiated) of the 
          planet from the start of the simulation is shown as a 
          function of time for the runs which produced Jupiter mass 
          planets. Most of the energy is emitted during the epoch
          when the planet accretes the bulk of its gas. This epoch
          is very similar (albeit displaced somewhat in time) for 
          runs 1G, 1sG, 1xsG and 2\textit{h}J, so the total energy
          radiated in these runs is almost the same. Accretion of
          gas near the luminosity peak is slower in the group 3 runs
          because the surface density is lower (and, in some cases,
          the viscosity is also lower); this allows the planet
          to radiate more energy.
          Diamonds denote bifurcation times.
          }
          \label{fig:ep}
\end{figure*}
\begin{figure*}[th!]
 \centering
\resizebox{0.65\hsize}{!}{\includegraphics[angle=90]{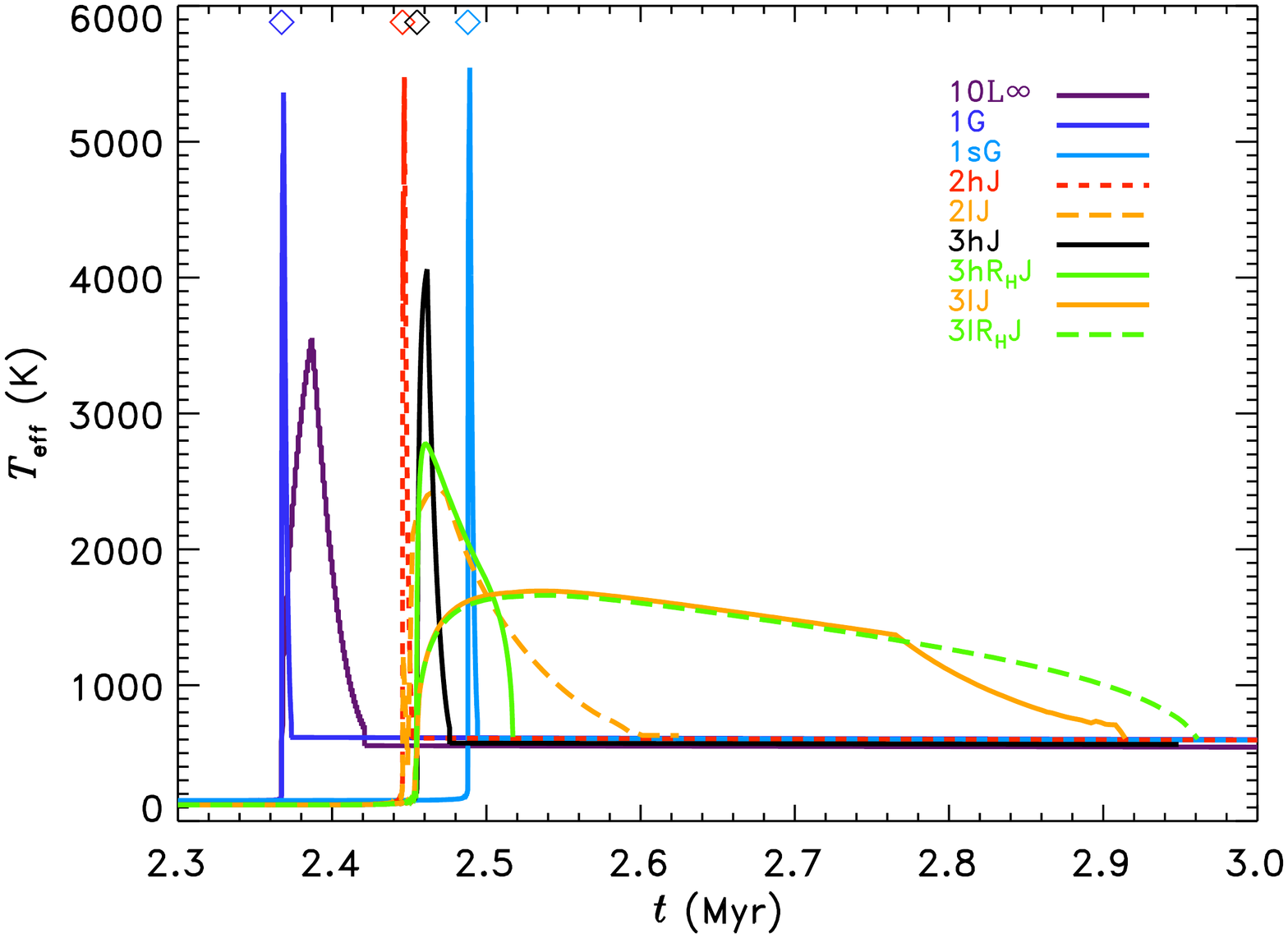}}
 \caption{The effective temperature of the planet during and 
          subsequent to the late phases of accretion for the runs 
          indicated. Note that in the most realistic runs, for which 
          accretion tapers off gradually, the planet has a lower peak 
          temperature, but it remains quite warm for longer 
          periods of time.
          The diamonds near the top left denote the bifurcation times.
          }
          \label{fig:teff}
\end{figure*}
Let us first examine the values for the quantities that were key 
findings of  previous simulations 
\citepalias{pollack1996,bodenheimer2000b,hubickyj2005}.  
These studies focused on the time required for the growth of the 
planet and the ultimate amount of condensable material in the 
planet. The crossover mass is almost identical in all of our
present runs (Table~\ref{table:2}, Fig.~\ref{fig:mpe}), 
as expected since in our formulation the crossover mass depends 
almost entirely on the surface density of 
solids, the mass of the star and the distance of the planet from 
the star \citepalias[see Eqs.~14 and 18 of][]{pollack1996}.
The value that we find for $M_Z$ at crossover, 
$16\,\mathrm{M}_{\oplus}$, is consistent with 
recent calculations of the mass of Jupiter's core based 
upon Galileo data \citep{militzer2008}; 
however other models yield smaller or no cores
\citep{saumon2004}.
The time at which crossover is reached is about $10^5$ years later
in Run~1s/1sG than in 10\Linf/1G
as a result of the smaller planet size (and thus smaller radiating
area, Table~\ref{table:2}); note that the planet's size is reduced 
by a larger factor late in the growth epoch. Crossover time is 
$3\times10^5$ years later in Run~1xsG than it is in 10\Linf/1G,
because the size of the 
planet is significantly smaller throughout accretion 
(Table~\ref{table:2}). 
Thus, \textit{reducing the planetary envelope's outer radius by a 
factor of four has a nontrivial affect on the time that it takes 
for the planet to reach crossover, but accretion rates are not 
so sensitive to this parameter for our remaining uncertainty to be 
significant}.
Crossover occurs $8\times10^4$ years sooner in Runs~2 than in 
Runs~1s, implying that the lower disk temperature, which
increases both the accretion radius,
$R_A$, and the gas density at the planet's outer boundary,
has a noticeable but not large effect on the growth rate.  
We found that this time was not very sensitive to gas density in
\citetalias{bodenheimer2000b} (compare Runs~P1 and P2, wherein 
gas densities differ by three orders of magnitude), but the 
sensitivity of the formation time to size of the planet's outer 
envelope boundary has not been previously studied.

Now let us turn to the growth of the planet at later times. 
These results are listed in Tables ~\ref{table:3} and \ref{table:4}, 
and the masses and radii of the planet at late times are plotted in 
Fig.~\ref{fig:mpl} and \ref{fig:rpl}, respectively.
In all cases, the planet's thermally regulated gas accretion
becomes so rapid that flow from the disk limits the planet's growth 
rate $\sim 10^{5}$ years subsequent to crossover.
The gas accretion rate of a planet with 
$M_p \gtrsim 50\,\mathrm{M}_{\oplus}$ that is well supplied by
the disk increases
so rapidly that, within each of the four groups that
bifurcated, the time at which disk-limited accretion set in 
varied by $\lesssim 2000$ years even though the value of
$\dot{M}_{XY}$ at this milestone differed by more than an order
of magnitude in some cases.

In our previous studies, the gas density in the disk remained 
constant and the ability of a planet to accrete gas was a 
non-decreasing function of the planet's mass, so we needed to 
terminate accretion in a highly artificial manner.  
While accretion in many of the runs presented herein was 
terminated when the planet reached Jupiter's mass, in some runs 
we allowed the planet to accrete until the gas density in its 
vicinity dropped to zero after $\le 3\,\mathrm{Myr}$ 
had elapsed.

Runs~2\textit{h} and 2\textit{l} were stopped abruptly at 
$3\,\mathrm{Myr}$,
yielding planets of mass $1830\,\mathrm{M}_{\oplus}$ 
($5.76\,\mathrm{M}_J$) and $525\,\mathrm{M}_{\oplus}$ 
($1.65\,\mathrm{M}_J$), respectively.
In Runs~3\textit{h} and 3\textit{l}, we assumed
that removal of gas by the planet's accretion does not affect 
the gas surface density of the solar nebula in which the planet 
is embedded.
As a result, planets of mass $1092\,\mathrm{M}_{\oplus}$ 
($3.44\,\mathrm{M}_J$) and $324\,\mathrm{M}_{\oplus}$
($1.02\,\mathrm{M}_J$) form by the time that the prescribed 
gas surface density of the nebula dropped to zero at 
$3\,\mathrm{Myr}$.

In half of the runs in group 3 (those whose designation includes
``R\sbs{H}''), the surface density of gas in the planet's accretion
zone drops as gas is accreted by the planet.
During most of the formation epoch, the planet's accretion rate 
of gas is smaller than the removal of gas from its feeding zone 
by the overall depletion of our protoplanetary disk.  
Thus, planetary removal of gas in Runs~3\textit{h}4R$_\mathrm{H}$ 
and 3\textit{h}R$_\mathrm{H}$J does not affect the planet's 
growth in our algorithm during this interval. 
Indeed, for the first $2.437\,\mathrm{Myr}$, one numerical
calculation suffices to follow all eight runs in group 3. 
However, when the planet begins to rapidly accrete gas, the 
surface density of gas within the planet's feeding zone drops 
rapidly, despite the addition of gas via the expansion of the 
feeding zone. This leads to a decline in $\dot{M}_{XY}$.  
Eventually, the accretion of gas by the planet becomes so 
small that the rate of decrease in gas surface density in the 
planet's accretion zone is less than the prescribed overall 
linear decline rate.  
At this point, we switch back to the linear drop-off (i.e., 
$\dot{M}_{XY}$ within the feeding zone being given by the 
derivative of Eq.~(\ref{eq:sig}) multiplied by the area of 
the feeding zone, plus a small addition of gas as a result of 
feeding zone expansion into regions of the disk that have not 
been depleted of gas via the planet's accretion).  
The run ends when the gas density in the feeding zone drops to 
zero.
In Run~3\textit{h}4R$_\mathrm{H}$, the planet grows to 
$403\,\mathrm{M}_{\oplus}$ ($1.27\,\mathrm{M}_J$) 
in $2.549\,\mathrm{Myr}$
and in Run~3\textit{l}4R\sbs{H} the planet reaches
$263\,\mathrm{M}_{\oplus}$ ($0.83\,\mathrm{M}_J$) 
in $2.825\,\mathrm{Myr}$.
In Run~3\textit{h}R$_\mathrm{H}$J, we reduce the size of 
the gas accretion zone to  $3.2\,R_\mathrm{H}$ 
so that the planet runs out of gas when its mass is 
$\sim 1\,\mathrm{M}_J$.
Likewise, in Run~3\textit{l}R\sbs{H}J we increase the size of the 
zone to $11\,R_\mathrm{H}$ to give the same final mass.

The planet radiates away much of the gravitational energy released 
by accretion and contraction, and therefore it is quite luminous 
during most of the accretionary epoch. 
The planet's luminosity throughout its growth is shown in 
Fig.~\ref{fig:lp} (top) for five cases which use the high viscosity disk 
formula to specify the hydrodynamic limit to gas accretion and which 
end with a Jupiter-mass planet. Peak luminosity values (see also 
Tables~\ref{table:3} and \ref{table:4}) are very similar for all 
of these runs apart from 3\textit{h}J, which has a smaller peak because 
the surface density of gas of the disk is much lower during the 
late stages of planetary accretion. Figure~\ref{fig:lp} (bottom) displays 
the planet's luminosity after bifurcation for selected runs that
ended with a $1\,\mathrm{M}_J$ planet. 
The luminosity peak in run 1G is higher but narrower than 
in 10\Linf\ because the specified hydrodynamic limit to accretion is 
larger. Likewise, the luminosity peaks for the 3\textit{h} runs are 
higher and narrower than those of the 3\textit{l} runs because of more 
rapid gas accretion. Slower accretion is also responsible for the 
slightly lower luminosity peaks for the `R\sbs{H}' runs compared to 
the runs with the same viscosity that do not account for disk depletion 
resulting from planetary accretion.

Note that the peak luminosity can be very high, up to 
$10^{-1.5}\,\mathrm{L}_{\odot}$ in Run~1G, but the width of the peak is 
very short in runs where the peak is high. 
During the planet's high luminosity phase, it heats the gas in its 
neighborhood of the circumstellar disk. However, as hydrodynamic 
factors (rather than thermal ones) limit the planet's accretion at this 
epoch, $\dot{M}_{XY}$ should not be significantly reduced. 
The total energy liberated during the rapid gas accretion 
phase is about the same in all cases (Fig.~\ref{fig:ep}), so in
Run~3\textit{l}R\sbs{H}J, for example, the peak luminosity is 
$10^{-3.5}\,\mathrm{L}_{\odot}$, but the phase lasts for almost 
$4\times 10^5$ years.

The integrated luminosity of the planet up to the time of crossover in 
every case is $\sim 10^{41}$ ergs. The integrated luminosity for the 
entire accretionary epoch for all cases that produce a $1\,\mathrm{M}_J$ 
planet is $3$ -- $4 \times 10^{42}$ ergs, with most of this being radiated 
during the (in some cases very) brief accretionary spike. 
The final gravitational potential energy of each of these $1\,\mathrm{M}_J$ 
planets is close to $-1.5 \times 10^{43}$ ergs. The integrated luminosity 
is significantly less than half the absolute value of the gravitational 
potential energy because dissociation of hydrogen absorbs a substantial 
amount of energy; the virial theorem for self-gravitating bodies is not 
violated. 

The effective temperature of a forming Jupiter-like planet is of interest 
to observers searching from thermal radiation of such objects. We show 
$T_{\mathrm{eff}} \equiv [L/(4\pi R^2_p \sigma_{\mathrm{rad}})]^{1/4}$ 
for several cases in Figure~\ref{fig:teff}.

\section{Implications for Capture of Irregular Satellites}
\label{sec:satellites}
\begin{figure*}[th!]
 \centering
\resizebox{0.65\hsize}{!}{\includegraphics[angle=90]{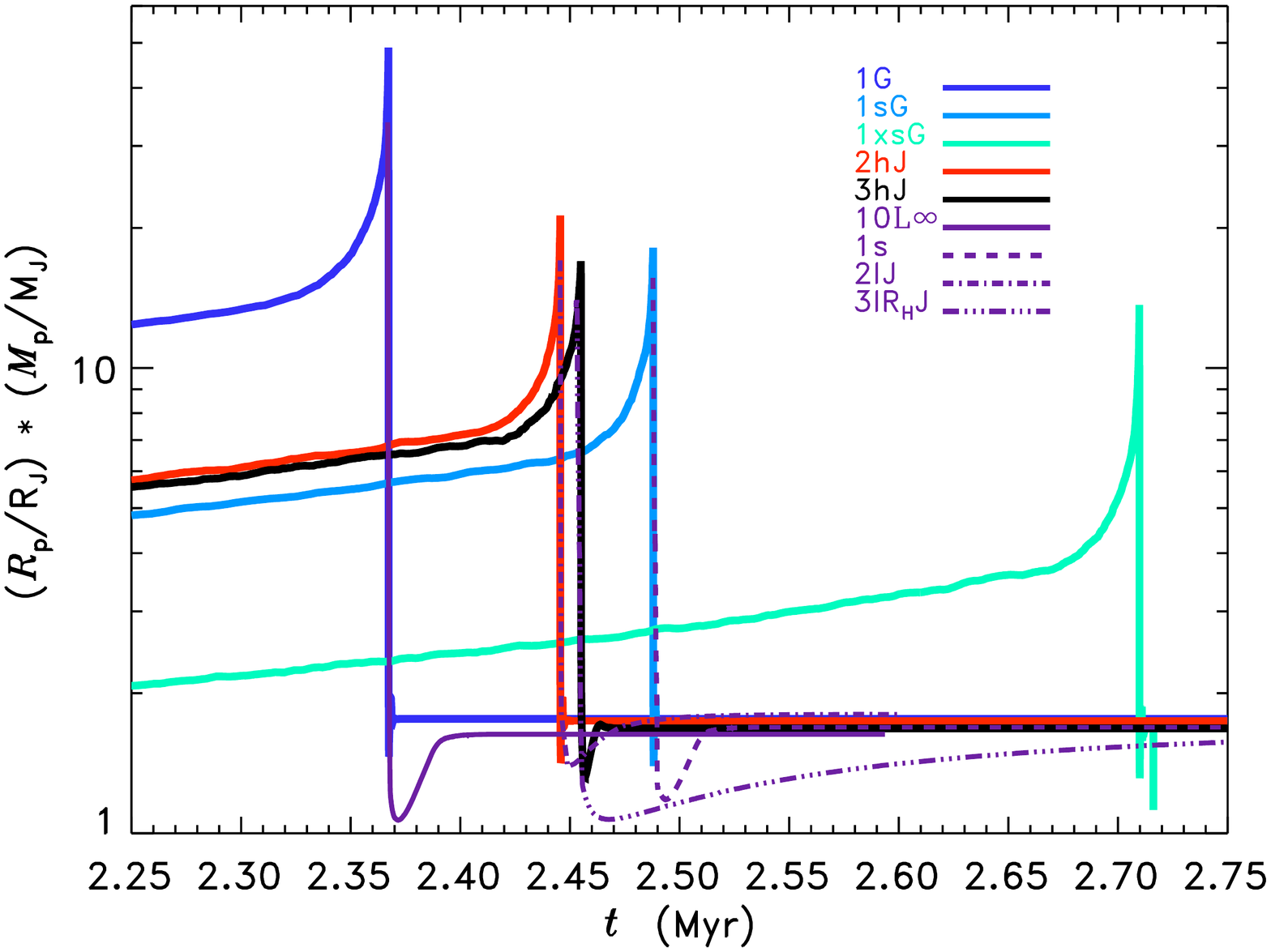}}
\resizebox{0.65\hsize}{!}{\includegraphics[angle=90]{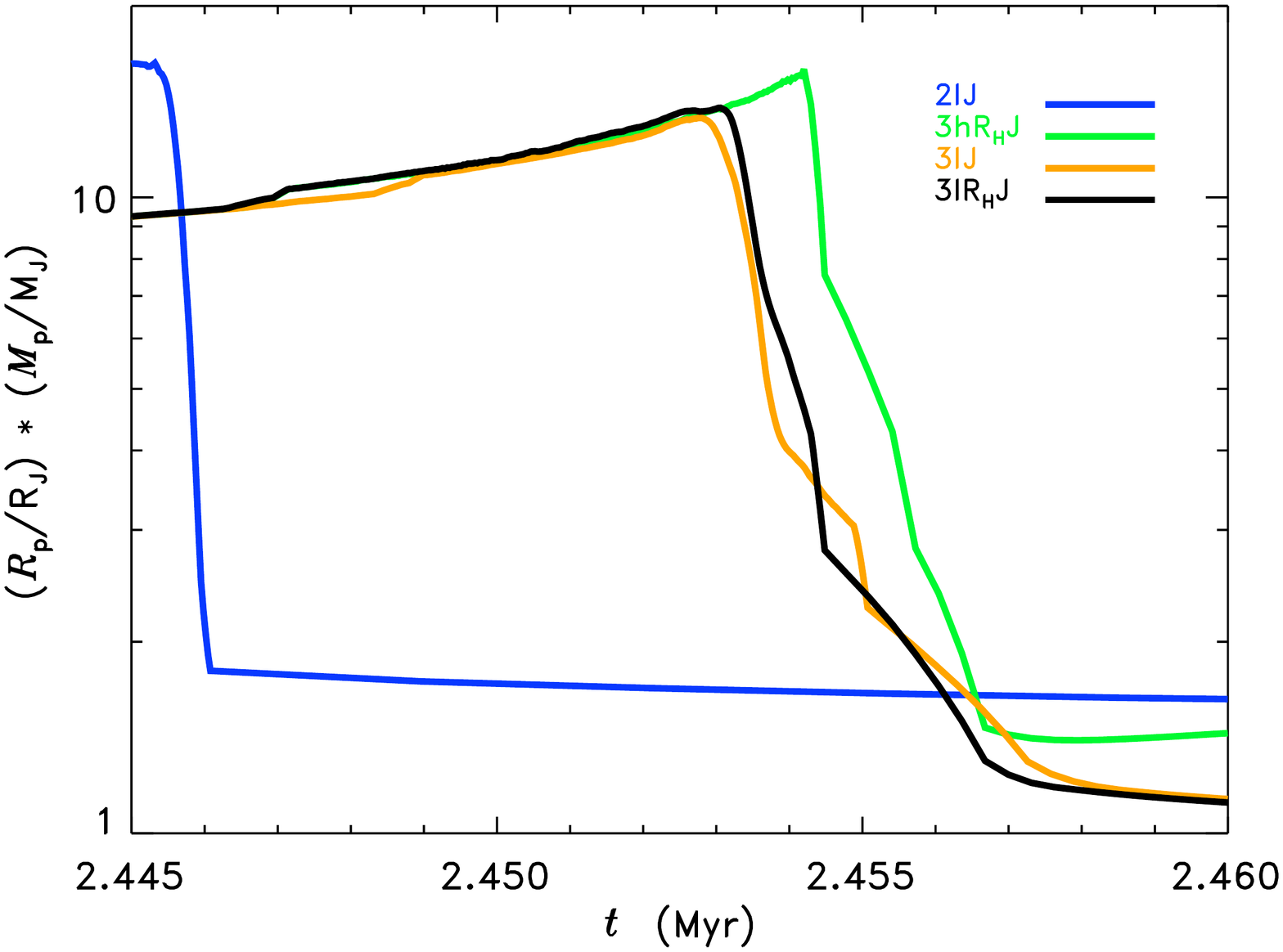}}
 \caption{The final orbital distance that a test particle orbiting 
          at the outer boundary of the planet at the time specified 
          would have at the end of the accretion epoch if the only 
          orbit-altering effect was shrinkage to conserve orbital
          angular momentum as the planet grew to a mass of 
          $1\,\mathrm{M}_J$. 
          Top: Late Phase~II and Phase~III for each 
          of the five groups, with two variants shown for 
          three of the four groups that bifurcated. Apart from 
          Runs~10\Linf\ and 1G, which we view to be among the least
          realistic
          runs of this paper (because of the large assumed size 
          of the planet and high disk viscosity), the peak value is 
          $\lesssim 20\,\mathrm{R}_J$. 
          Note that all of the curves end at 
          values of $\sim 1.7\,\mathrm{R}_J$, because in our model
          $1\,\mathrm{M}_J$ planets are about this radius at the
          termination of accretion (Tables~\ref{table:3} and 
          \ref{table:4}). 
          Bottom: Close-up values near the peak for the four most
          physically realistic runs presented in this paper. The
          characteristic shrinking times range from a few hundred
          to a few thousand years.
          }
          \label{fig:srp}
\end{figure*}
The irregular satellites of the giant planets are clearly captured
objects \citep[e.g.,][]{peale2007}. 
The semimajor axes of the orbits of all but $2$ of the $54$ known 
jovian irregular satellites are between
$150$ and $350\,\mathrm{R}_J$; the orbital periods of these objects
range from $8$ months to a little over $2$ years. 
An object approaching from 
heliocentric orbit has positive energy with respect to Jupiter 
(i.e., the magnitude of their kinetic energy exceeds that of their 
gravitational binding energy to Jupiter). Three-body effects 
caused by the combined perturbations of the Sun and Jupiter near 
the boundary of Jupiter's Hill sphere can reverse this balance, 
leading to temporary capture into jovocentric orbit 
\citep[e.g.,][]{kary1996}. Permanent capture requires that a 
non-conservative process acts while the body is in the vicinity of 
Jupiter.

Addition of mass to Jupiter increases the depth of the planet's 
gravitational potential well. Jupiter's Hill sphere expands 
(see Eq.~\ref{eq:RH}), and orbiting objects conserve angular 
momentum as the planet accretes additional material. Assuming that 
the orbital period of a moon is short compared to the timescale of 
the increase of the planet's mass, eccentricity is conserved and 
the semimajor axis of the orbit goes as $M_p^{-1}$. Objects captured 
when Jupiter was significantly less massive than it is today thus 
would orbit well within the current planet's Hill sphere, because their 
immediate post-capture orbits would have been within the smaller 
Hill sphere of the growing planet and the orbits would have shrunk during 
Jupiter's later accretion of matter. (Exceptions would be objects 
captured into distant quasi-satellite orbits which fortuitously
crossed the narrow bridge in phase space connecting them with
retrograde orbits as the planet grew and those whose orbits 
were altered by collisions, etc.) 

No proposed mechanism for capture of small satellites by giant planets 
is fully satisfactory \citep{jewitt2007}. The viability of two of the 
leading theories for the capture of irregular satellites (apart from 
Neptune's anomalous large irregular moon, Triton) can be assessed 
within the framework of our model of giant planet growth. These 
capture theories involve: (i) reduction of kinetic energy via gas 
drag within primordial circumplanetary envelopes shortly before 
these envelopes collapsed \citep{pollack1979}; and (ii) increase in 
the magnitude of the gravitational potential energy via rapid mass 
accretion by Jupiter \citep{heppenheimer1977}. 

\citet{pollack1979}  used Bodenheimer's (\citeyear{bodenheimer1977}) 
model of the evolution of giant planet envelopes. According to 
Bodenheimer's model, Jupiter collapsed hydrodynamically from a size 
of $\sim 225\,\mathrm{R}_J$ to $\sim 30\,\mathrm{R}_J$ on a timescale 
of $\sim 1$--$2$ years, and this process occurred after the planet 
had reached its present mass, $1\,\mathrm{M}_J$. Planetesimals 
entering the envelope shortly before this collapse could have lost 
enough kinetic energy via gas drag to be captured, but not so much 
as to spiral too deeply into the planet's gravitational potential 
well. Our present model of giant planet formation differs 
substantially from that prevailing three decades ago, and we show 
below that capture of the observed irregular satellites by gas drag 
in the extended envelope of proto-Jupiter is not consistent with the 
planet growth simulations presented in Section~\ref{sec:results}.

The giant planet formation models presented herein have three major
differences from those of \citet{bodenheimer1977} that make irregular
satellite capture in the extended and thermally-supported envelope phase 
difficult. Firstly, the envelope collapses gradually, over a timescale
of a few centuries to a few millennia for cases (Fig.~\ref{fig:srp},
bottom) 
in which the most realistic assumptions of the rate that the disk 
supplies gas to the planet are made (and over decades in some of the 
less physically plausible runs) rather than hydrodynamically; this 
makes survival of captured satellites against spiraling into the 
planet questionable. Secondly, the planet's envelope collapses when 
the planet's mass is only about $0.25\,\mathrm{M}_J$ 
(Tables~\ref{table:3} and \ref{table:4});
so orbits of bound bodies would shrink substantially due to the
continued accretion of mass by the planet.
Thirdly the size of the 
planet's envelope prior to collapse is only $\lesssim 80\,\mathrm{R}_J$ 
for all runs that use the prescription for planet radius given by the 
hydrodynamic simulations presented in Section~\ref{sec:envsize}. 

The latter two differences imply that satellites captured during
Jupiter's extended and thermally-supported envelope phase would orbit 
far closer to the planets than do the observed irregulars. The maximum
size of the envelope (which is the size prior to collapse) is smaller 
than calculated by \citet{bodenheimer1977} both because the envelope
collapses at smaller planetary mass and because in our new models
the envelope is restricted by gas flow in the protoplanetary disk
to a region only $\sim 1/4$th as large as that of their Hill radius
(Fig.~\ref{fig:boundtracers} and \ref{fig:tracers_xy}). We plot
in Fig.~\ref{fig:srp} the temporal evolution of a `scaled radius', 
$\mathcal{R}_p \equiv R_p\, M_p/\mathrm{M}_J$, which is the 
post-accretion perijove distance of a body on an orbit just tangent 
to the outer radius of the protoplanet at the time in
question\footnote{This assumes that the accreting gas does not collide
with the satellites, leading to gas drag and further shrinking satellite
orbits, and that accretion is slow compared to the orbital period of 
the satellites. The most 
rapid rate of increase in Jupiter's mass depends upon the
viscosity in the surrounding protoplanetary disk (Fig.~\ref{fig:Mdot_xy}
and Table~\ref{table:3}). For the more viscous case that we studied,
the planet's mass increases at up to $\sim 0.3$\% per year.}.  
The value of $\mathcal{R}_p$ peaks below $50\,\mathrm{R}_J$ in all of 
our runs and below $20\,\mathrm{R}_J$ in the most realistic runs 
(Fig.~\ref{fig:srp}, bottom).

The simulations of the growth of Jupiter presented herein are thus 
inconsistent with the model of capture of Jupiter's irregular satellites 
within proto-Jupiter's distended and thermally-supported envelope.  
Our calculations do not address (and therefore do not exclude) the 
possibility that the irregular satellites were captured as a result of 
gas drag within a circumjovian disk. 
Additionally, 
the accretion timescale during the epoch when the planet accumulates
the bulk of its mass is short compared to the planet's overall growth
time, yet too long compared to trapping times for temporary capture 
of particles by Jupiter for the \citet{heppenheimer1977} model to be 
viable.

\section{Summary and Conclusions}
\label{sec:summary}
\begin{figure}
\centering%
 \resizebox{1.0\hsize}{!}{\includegraphics[angle=90]{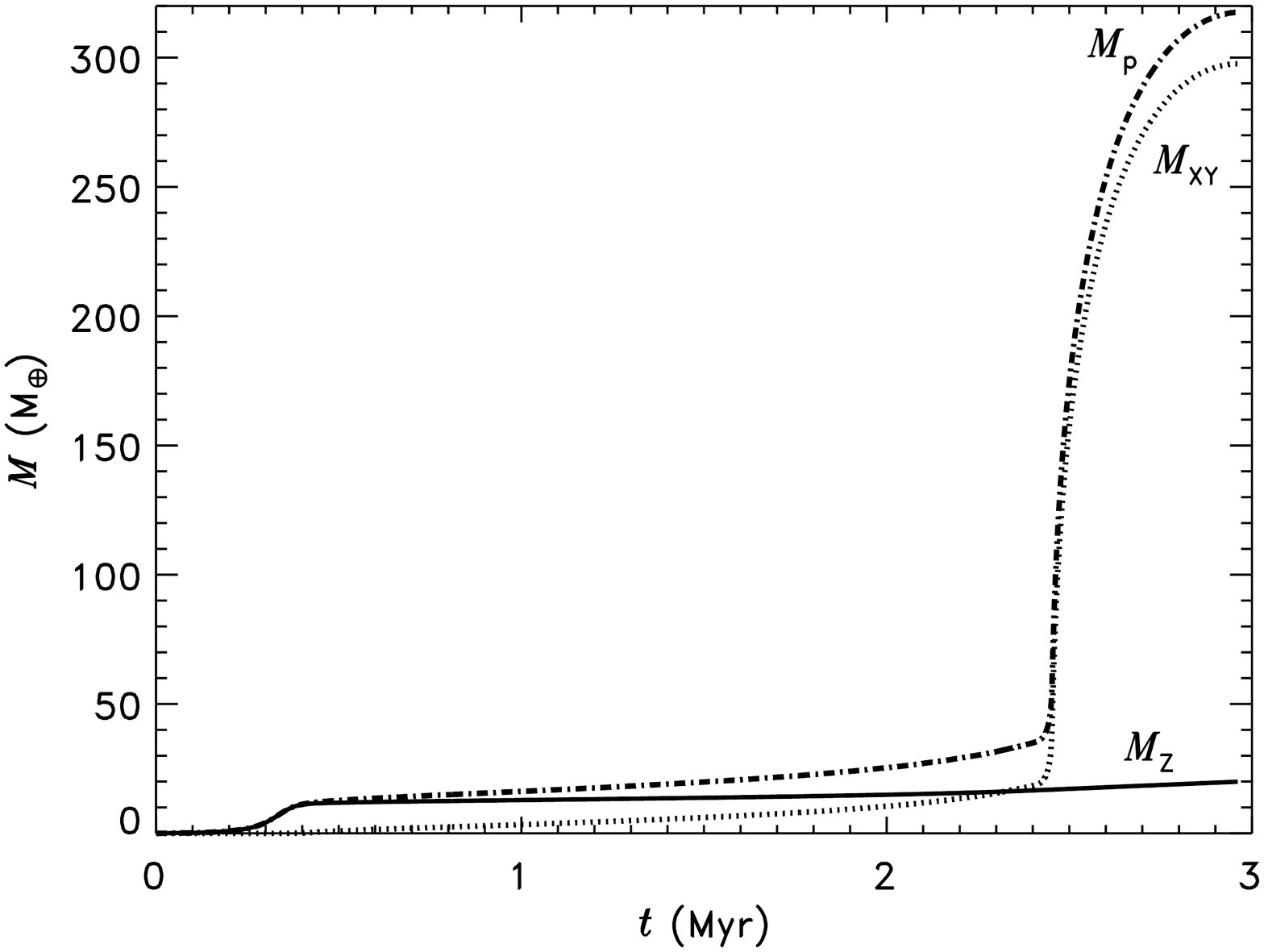}}
 \resizebox{1.0\hsize}{!}{\includegraphics[angle=90]{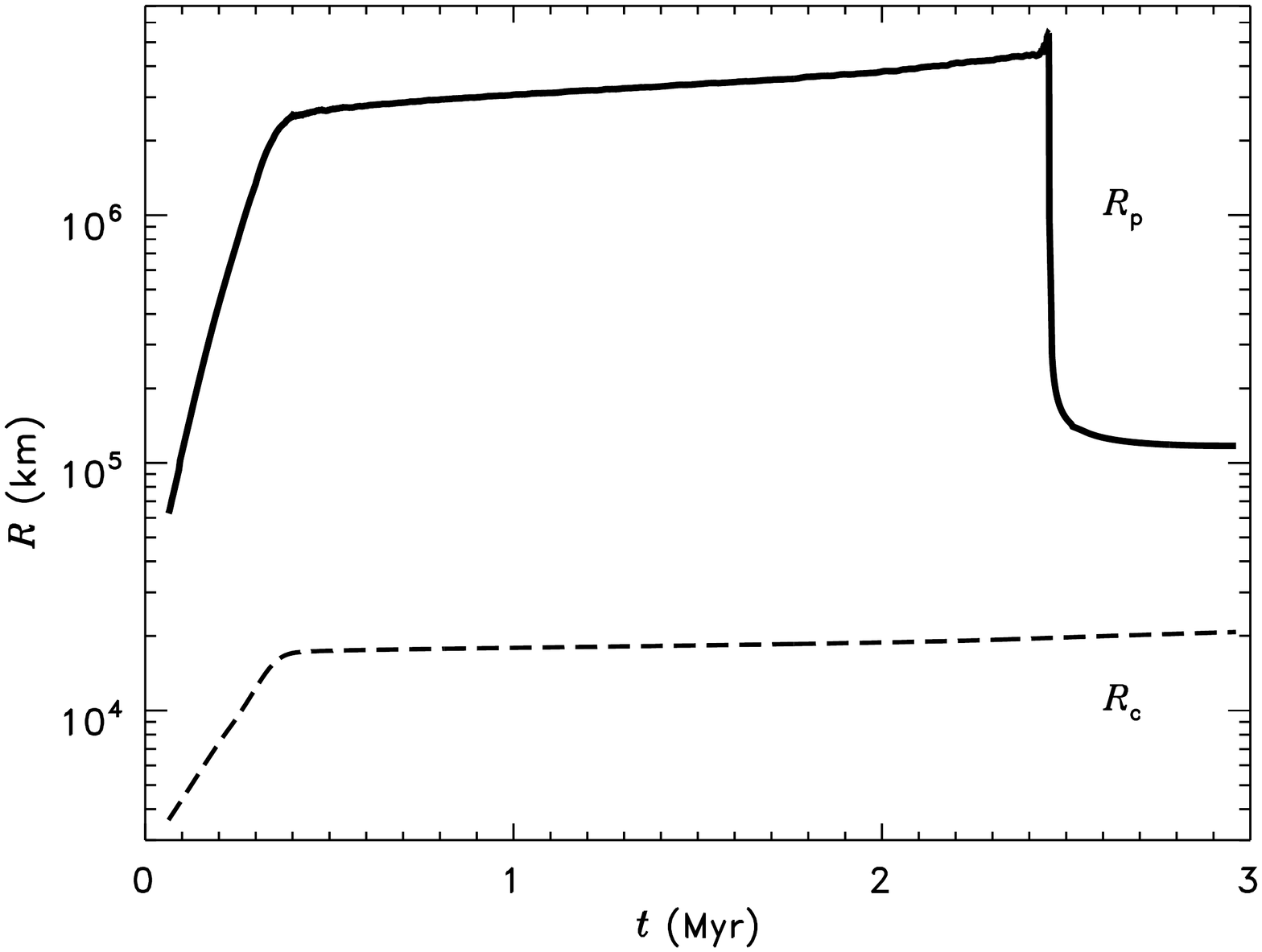}}
 \resizebox{1.0\hsize}{!}{\includegraphics[angle=90]{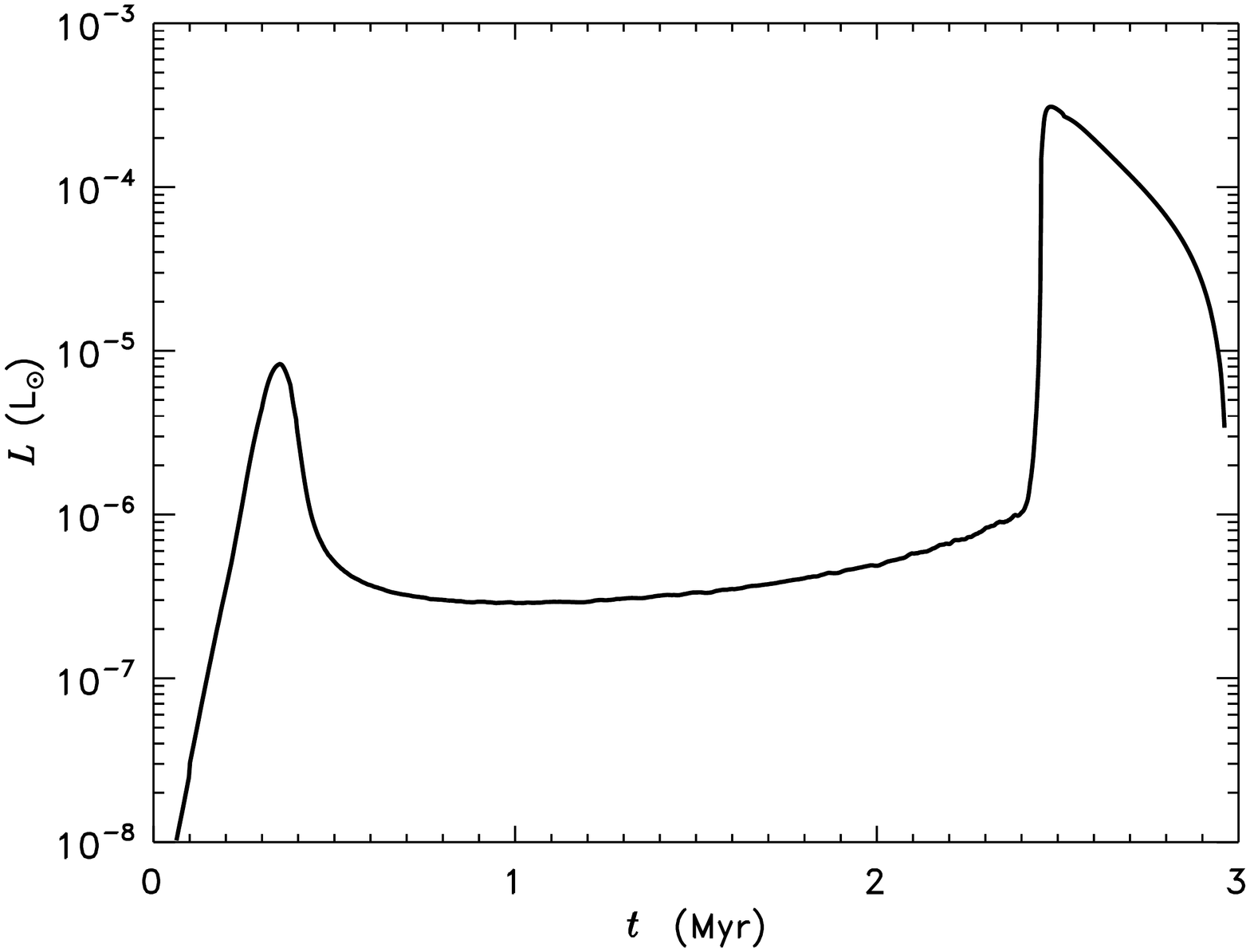}}
  \caption{The temporal evolution of the planet according to our most 
           physically realistic case, Run~3\textit{l}R\sbs{H}J (see
           Tables~\ref{table:1}, \ref{table:2}, and \ref{table:4} 
           for details). Top: The mass of 
           solids in the planet (solid line), gas in the planet 
           (dotted line) and 
           the total mass of the planet (dot-dashed line) are shown as 
           functions of time. Note the slow, gradually increasing, 
           buildup of gas, leading to a rapid growth spurt, followed by
           a slow tail off in accretion. Middle: The radius of the planet
           (solid line) and that of the planet's heavy element core
           (dashed
           line) are shown as functions of time. Note the logarithmic
           scale used for radius. Bottom: The planet's luminosity is shown 
           as a function of time. The rapid contraction of the 
           planet just before $t = 2.5\,\mathrm{Myr}$ coincides with its 
           highest luminosity and the epoch of most rapid gas accretion. 
           }
  \label{fig:prc}
\end{figure}
We have modeled the growth of Jupiter incorporating both thermal and
hydrodynamic constraints on its accretion of gas from the circumsolar
disk.
Our study included simulations of planets growing in disks of high and 
low viscosity; the surface mass density of gas within the disk remained 
constant in some cases and decreased gradually in others 
(see Table~\ref{table:1} for details). The most physically plausible 
model of the formation of Jupiter included herein is 
Run~3\textit{l}R\sbs{H}J, in which the dimensionless disk viscosity is  
$\alpha = 4 \times 10^{-4}$ and the surface density of the gas within 
the disk decreases with time, thereby allowing a gradual tapering off 
of gas accretion as the planet approaches its ultimate mass. Results 
of this simulation are presented in Figure~\ref{fig:prc}.

The principal results of our investigation are:
(1) Three dimensional hydrodynamic calculations show that the flow of
gas in the circumsolar disk limits the region occupied by the planet's
tenuous gaseous envelope to within $\sim 0.25\,R_{\mathrm{H}}$ (Hill
sphere radii) of the planet's center, which is much smaller than the 
value of  $\sim 1\,R_{\mathrm{H}}$ that was assumed in previous studies.
(2) This smaller size of the planet's envelope increases the planet's
accretion time, but only by $\sim 5$--$10$\%. In general, in agreement with
previous results \citepalias{hubickyj2005}, Jupiter formation times are
in the range $2.5$--$3\,\mathrm{Myr}$, assuming a protoplanetary disk
with solid surface density of $10\,\mathrm{g\,cm}^{-2}$ and dust opacity
in the protoplanet's envelope equal to $2$\% that of interstellar material.
(3) In a protoplanetary disk whose dimensionless viscosity parameter 
$\alpha \sim 4 \times 10^{-3}$, giant planets grow to several times
the mass of Jupiter unless the disk has a small local surface density
when the planet begins to accrete gas hydrodynamically, or the disk is
dispersed very soon thereafter. The large number of planets known with
masses near Jupiter compared with the smaller number of substantially
more massive planets \citep{udry2007} is more naturally explained by
planetary growth within circumstellar disks whose dimensionless viscosity
parameter is $\alpha \sim 4 \times 10^{-4}$.
(4) Capture of Jupiter's irregular satellites within the planet's diffuse
and distended thermally-supported envelope is not consistent with the Jupiter 
formation models presented herein.

\acknowledgments

We thank Jeff Cuzzi and two anonymous referees 
for providing valuable comments.
Primary support for this study was provided by NASA's Outer Planets 
Research Program grant 344-30-99-02;
additional support came from NASA Origins of Solar Systems 
grant NNX08AH82G.
GD is supported through the NASA Postdoctoral Program.
Computational resources for the 3-D hydrodynamic calculations were provided 
by the NASA High-End Computing Program systems under grants SMD-06-0181
and SMD-07-0372.

\end{document}